\documentclass[12pt,letterpaper]{article}
\pdfoutput=1
\usepackage{jheppub}
\usepackage[english]{babel}
\usepackage[latin1]{inputenc}
\usepackage{amsfonts}
\usepackage{amssymb,amsmath,mathabx}
\usepackage{graphicx}
\usepackage{framed}

\title{On the Amplitude/Wilson Loop duality in  $\mathcal{N}=2$ SCQCD}

\author[]{Marta Leoni$^{\ast, \hash}$,} 
\author[\dag]{Andrea Mauri}
\author[\hash]{and Alberto Santambrogio}

\affiliation[\ast]{Dipartimento di Fisica dell'Universit\`a degli studi di Milano, Via Celoria 16, I-20133 Milano, Italy}
\affiliation[\hash]{INFN, Sezione di Milano, Via Celoria 16, I-20133 Milano, Italy}
\affiliation[\dag]{Dipartimento di Fisica dell'Universit\`a degli studi di Milano--Bicocca, Piazza della Scienza 3, I-20126 Milano, Italy}

\emailAdd{marta.leoni@mi.infn.it}
\emailAdd{andrea.mauri@mi.infn.it}
\emailAdd{alberto.santambrogio@mi.infn.it} 

\abstract{We compute the four--point amplitude with external adjoint particles in $\mathcal{N}=2$ SCQCD at two loops using $\mathcal{N}=1$ superspace Feynman diagrams, extending the results of \href{http://arxiv.org/abs/1406.7283}{arXiv:1406.7283}. We consider the diagrammatic difference with the corresponding process of $\mathcal{N}=4$ SYM finding a non vanishing result, which is a non trivial function of the kinematic variables. This demonstrates that  in $\mathcal{N}=2$ SCQCD, even in the sector with external particles in the vector multiplet,  the amplitude/Wilson loop duality is inevitably broken at two loops.\\[0.7cm]
 \vspace{-0.3cm}
}

\preprint{February 2015 \\ \vspace{-1cm} \begin{flushright} IFUM-1036-FT \end{flushright}}


\allowdisplaybreaks[1]

\newcommand {\non}{\nonumber}
\newcommand{\Li}[2]{{\mbox{Li}}_{#1}\left(#2\right)}

\newcommand{\mc}{\mathcal}
\def\tr{\mathrm{Tr}}
\def\d{\mathrm{d}}

\def\Tr{\mathrm{Tr}}
\def\ln{\text{ln}}

\def\bseq{\begin{subequation}}  
\def\eseq{\end{subequation}}
\def\bsea{\begin{subeqnarray}}  
\def\esea{\end{subeqnarray}}

\newcommand{\beq}{\begin{equation}}
\newcommand{\bea}{\begin{eqnarray}}
\newcommand{\eea}{\end{eqnarray}}
\newcommand{\eeq}{\end{equation}}

\newcommand{\g}{\gamma}

\newcommand{\e}{\epsilon}

\newcommand{\p}{\pi}

\begin{document}
\maketitle 

\section{Introduction}

In this paper we pursue the analysis of \cite{Leoni:2014fja} by computing the two--loop amplitude with four adjoint external particles in $\mathcal{N}=2$ superconformal QCD. In \cite{Leoni:2014fja} we have computed the one--loop four--point scattering amplitudes with general external fields and the two--loop amplitude of four fundamental particles. The calculation we present in this paper  extends the two--loop analysis to the adjoint sector and assumes a special relevance since it provides a test of the presence of the duality between scattering amplitudes and light-like Wilson loops in $\mathcal{N}=2$ SCQCD. 
 
The original formulation of the duality was given in the planar $\mathcal{N}=4$ SYM theory where it was first introduced at strong coupling \cite{Alday:2007hr} and then found at weak coupling in \cite{Drummond:2007aua, Brandhuber:2007yx, Drummond:2007cf}. The duality relates to all orders in perturbation theory the divergent and the finite parts of MHV $n$--point scattering amplitudes to the expectation value of light--like polygonal Wilson loops. Its presence is connected to the existence of a hidden dynamical symmetry of the amplitudes, the so called dual conformal symmetry, which is obscured by the off--shell Lagrangian formulation of the model (see e.g. \cite{Elvang:2013cua} for a recent review of the subject). The generators of the dual conformal symmetry and of the standard conformal symmetry close into an infinite dimensional Yangian algebra \cite{Drummond:2009fd}. The presence of the Yangian algebra is believed to be the manifestation of the integrability of planar $\mathcal{N}=4$ SYM in the context of scattering amplitudes. 

At the moment it is not clear which class of models should display such symmetries besides $\mathcal{N}=4$ SYM. In the three--dimensional ABJM theory \cite{Aharony:2008ug} the planar four--point amplitude has been found to be dual conformal invariant and to coincide with the light--like four--sided Wilson loop up to two loops \cite{Agarwal:2008pu,Henn:2010ps,Chen:2011vv, Bianchi:2011dg}, suggesting that the amplitude/Wilson loop duality is valid also in this theory. 
Outside the four--point case the duality is not expected to have the standard form due to the fact that the ABJM amplitudes are not MHV, except for the four--point one. Nevertheless the six--point amplitude has been shown to be dual conformal invariant up to two loops \cite{Bargheer:2012cp, Bianchi:2012cq, CaronHuot:2012hr} and the hexagonal light--like Wilson loop was computed up to two loops \cite{Henn:2010ps,Bianchi:2011rn,Bae:2014xca}.

The aim of this paper is to check whether the dual conformal symmetry and the scattering amplitude/Wilson loop duality are present in $\mathcal{N}=2$ SCQCD. 
The computation of the four--sided Wilson loop in $\mathcal{N}=2$ SCQCD was already performed up to three loops \cite{Andree:2010na}, finding that the $\mathcal{N}=4$ SYM and the $\mathcal{N}=2$ SCQCD results match up to two loops while they are different at three loops.
In \cite{Leoni:2014fja} we confirmed the result first derived in \cite{Glover:2008tu}, by computing with $\mathcal{N}=1$ super Feynman diagrams  the corresponding four--point one--loop amplitude and finding that the duality is valid at one--loop order. In this paper we tackle the problem at two--loop order. This computation is a crucial test for the presence of the duality: if the two--loop amplitude is equal to the one of  $\mathcal{N}=4$ SYM up to irrelevant constants the duality is valid also at two loops, while if it is different the duality is broken. Our main result is given in equation (\ref{result}), where
we show that the difference between the two--loop amplitude in $\mathcal{N}=2$ SCQCD and the analogous one in $\mathcal{N}=4$ SYM consists of a non vanishing expression, which is a non trivial function of the kinematic variables. The form of our result implies that the scattering amplitude/Wilson loop duality is not valid in $\mathcal{N}=2$ SCQCD. We also find that  the maximum transcendentality principle  \cite{Kotikov:2001sc} is violated at two--loop order in the adjoint sector. 

The paper is organized as follows: in Section \ref{sec2} we review $\mathcal{N}=2$ SCQCD and the known results concerning scattering amplitudes. In Section \ref{sec3} we present the computation of the two-loop amplitude with four external particles in the adjoint representation of the gauge group. In  Section \ref{sec4} we comment our result and then we conclude.

\section{Scattering in $\mathcal{N}=2$ SCQCD}\label{sec2}

We consider the $\mathcal{N}=1$ superfield formulation of $\mathcal{N}=2$ SCQCD introduced in \cite{Leoni:2014fja}, which is based on the four--dimensional superspace notations of \cite{Gates:1983nr}. The superspace action in the Fermi--Feynman gauge is given by 
\begin{align}\label{action}
S&=   \,\, S_0 + S_{gf} \non \\
S_0 & =  \int\d^4x \d^4\theta \bigg[  \tr\big(e^{-g V} \bar{\Phi}e^{g V}\Phi\big) + \bar{Q}^{ I}e^{g V}Q_{ I}  + \tilde{Q}^{ I}e^{-g V}\bar{\tilde{Q}}_{ I} \bigg]\,\,   +  \\
& + \frac{1}{g^2} \int\d^4x\d^2\theta \   \tr\big(W^{\alpha}W_{\alpha}\big) ++i g \int\d^4x\d^2\theta  \ \tilde{Q}^{ I} \Phi Q_{ I}
-i g \int\d^4x\d^2\bar\theta  \  \bar Q^{ I}\bar\Phi \bar{\tilde{Q}}_{ I}  \non \\
S_{gf} & =  \int d^4x d^4\theta \ \Tr\left(- (D^2V) (\bar{D}^2V) +(c'+\bar{c}')\textrm{L}_{\frac{g V}{2}}\big[c+\bar{c}+\coth \textrm{L}_{\frac{g V}{2}}(c-\bar{c}) \big]\right) \non  
\end{align}
This model is a $SU(N)$ gauge theory described in terms of an adjoint vector superfield $V$ with  superfield strength $W_\alpha = i \bar{D}^2(e^{-g V}D_{\alpha}e^{g V})$. The ghosts are introduced by anticommuting chiral superfields $c$ and $c'$ and are coupled to the vector superfield $V$ through $\textrm{L}_{\frac{g V}{2}} X = \frac{g}{2}[V,X]$.  The matter is described by an adjoint chiral superfield $\Phi$ and by a pair of chiral superfields $Q_I$ and $\tilde{Q}^I$, transforming respectively in the fundamental and antifundamental representation of the gauge group $SU(N)$. The fundamental fields carry an additional $U(N_f)$ flavour index $I=1,\dots,N_f$. The number of flavours is fixed by the following condition: $N_f=2N$, which assures that the model is conformal invariant. 

We compute four--point amplitudes in perturbation theory considering the planar Veneziano limit, which consists in taking $N$ and $N_f$ large, keeping their ratio fixed. We perform the calculations computing all $\mc N=1$ super Feynman diagrams which contribute to a selected superamplitude, and then we consider only the projection of the superamplitude on the four--scalar component.  We refer the reader to \cite{Leoni:2014fja} for more details on superspace conventions, Feynman rules and a step by step description of the computational techniques.

The  four--point amplitudes can be classified into three independent sectors: the adjoint, the mixed and the fundamental one, with four, two and zero adjoint external superfields respectively. Inside each sector, amplitudes are related by supersymmetry transformations. In \cite{Leoni:2014fja} the one--loop amplitudes were analytically computed in all of the three sectors. In the mixed and fundamental sectors the one--loop amplitudes lose the dual conformal invariance but nevertheless they still exhibit maximum transcendentality weights. In the fundamental sector the computation was pushed up to two loops, showing that at this order both dual conformal invariance and maximum transcendentality are not present.

In the next Section we will present the two--loop computation of the four--point amplitude in the adjoint sector. In order for the amplitude/Wilson loop duality to be present we should find a vanishing difference between the $\mathcal{N}=2$ SCQCD amplitude and the $\mathcal{N}=4$ SYM one. In the next Section we will show that this is not the case.

\section{The two--loop adjoint amplitude}\label{sec3}

We now present the computation of the subamplitude $\mc A^{(2)}(\Phi(1) \bar{\Phi}(2) \bar{\Phi}(3) \Phi(4))$ at two loops for the process $(\Phi \bar{\Phi} \bar{\Phi} \Phi)$ in $\mathcal{N}=2$ SCQCD. 

A convenient way to perform this calculation is to consider the diagrammatic difference with the $\mc N=4$ SYM process with four external adjoint superfields with equal flavour indices, namely $(\Phi_1 \bar{\Phi}_1 \bar{\Phi}_1\Phi_1)$. Doing a projection, it is possible to extract from the $\mathcal{N}=4$ SYM superamplitude the component which corresponds to the MHV gluon amplitude, whose two--loop correction was found long ago using unitarity cuts \cite{Bern:1997nh}.  Therefore, instead of computing from scratch the whole two--loop superamplitude in $\mathcal{N}=2$ SCQCD  we consider just the diagrammatic difference between the above processes in the two models. In this way we are left with a manageable number of diagrams that have to be computed directly, while the majority of the diagrams gets canceled because they give identical expressions.  A similar reasoning was first used in \cite{Andree:2010na} to compute the difference between closed Wilson loops in $\mathcal{N}=2$ SCQCD and $\mathcal{N}=4$ SYM theory.

The first step in our computation is to identify which diagrams contributing to the two--loop amplitude might yield different results in $\mathcal{N}=4$ SYM and in $\mathcal{N}=2$ SCQCD.  The only way to obtain diagram topologies which produce    {\it a priori}  different contributions in the two models is to draw chiral lines which admit a realization in terms of fundamental fields in $\mathcal{N}=2$ SCQCD \footnote{For general external flavour configurations in $\mathcal{N}=4$ SYM   there is an exception to this rule, which is not present in the case of equal flavour external legs we are considering here.}. Moreover, since we are considering  an amplitude  with adjoint external fields, the fundamental lines of the $\mathcal{N}=2$ SCQCD diagrams must inevitably form closed circuits. We already stated that diagrams containing loops of fundamental fields in $\mathcal{N}=2$ SCQCD give the same results of the corresponding diagrams in $\mathcal{N}=4$ SYM with loops of adjoint fields with different flavours.  Nevertheless, following the reasoning of \cite{Pomoni:2011jj, Pomoni:2013poa} and specifying it to the special case of $\mathcal{N}=2$ SCQCD, it is possible to show that  whenever the chiral circuits are cut by an adjoint internal line they give rise to a non vanishing difference. These diagram topologies turn out to be the only ones which give a contribution. In the rest of this Section  we support with explicit examples these statements. Presenting our computation we consider only diagrams containing chiral loops cut by an adjoint internal line, but as a test of the arguments expressed above we have explicitly checked case by case that all the other two--loop diagrams do not contribute to the difference of the amplitudes.  

Eventually, we organize the diagrams which contribute to the diagrammatic difference in the following classes: two--loop vertex and propagator corrections to the tree--level diagram and genuine two--loop diagrams. We now study each classes separately.

\subsection{Propagator corrections} \label{propsec}

\begin{figure} [ht]
\centering
 \includegraphics[scale=0.6]{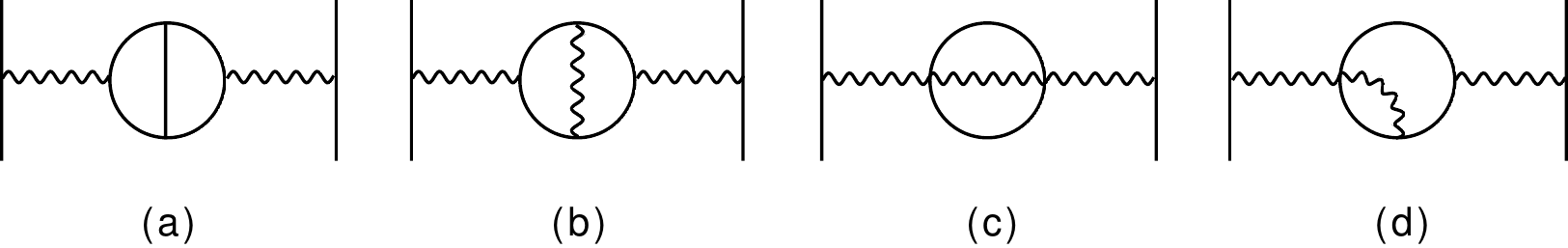} 
\caption{Diagram topologies containing the two--loop propagator correction which give a non vanishing contribution to the diagrammatic difference.} \label{proptot} 
\end{figure}

The first class of diagrams we consider is given by the two--loop dressings of the gauge superfield propagator in the tree--level process.  The diagram topologies giving a non vanishing contribution to the amplitude difference are listed in Fig. \ref{proptot}. \vspace{0.2cm} 

\begin{figure} [ht]
\hspace{0.5cm}
 \begin{minipage}{4cm}
\includegraphics[scale=0.35]{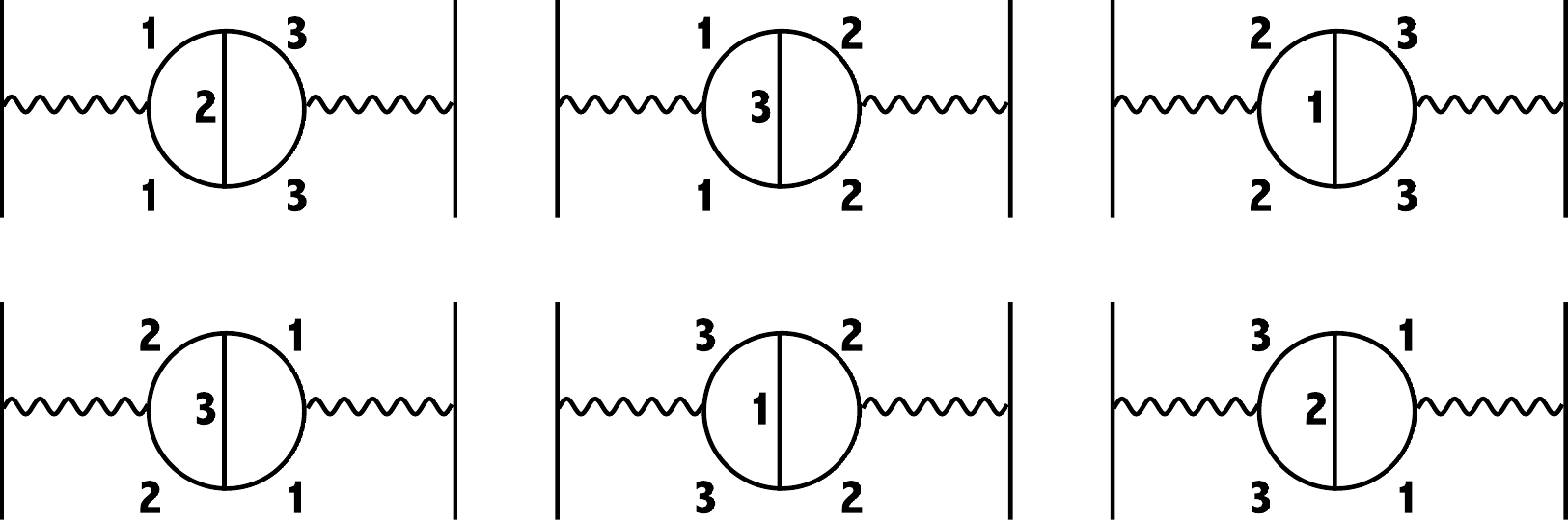}
\end{minipage}  
\hspace{2.7cm}
\begin{minipage}{4cm}
\includegraphics[scale=0.35]{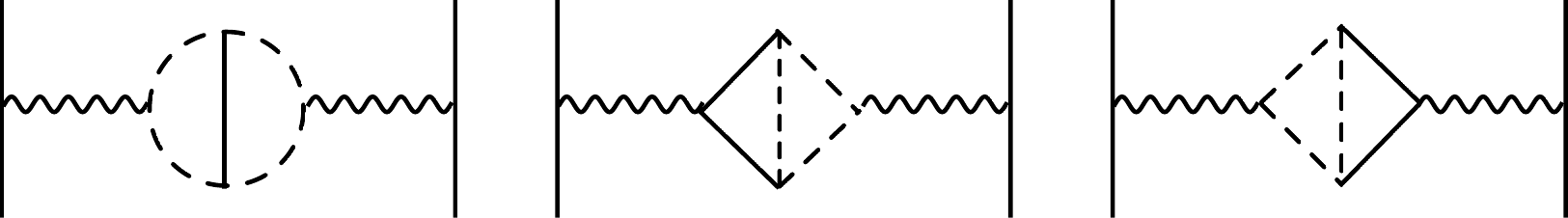}
\end{minipage}  \vspace{0.3cm}
\caption{$\mathcal{N}=4$ SYM diagrams of topology (a) on the left side and $\mathcal{N}=2$ SCQCD ones on the right side.}  \label{aprop}
\end{figure}

In $\mathcal{N}=4$ SYM the diagram (a) of Fig. \ref{proptot} can be drawn with six different flavour flows, as shown in the left side of Fig. \ref{aprop}, all giving the same contribution. There are three diagrams in $\mathcal{N}=2$ SCQCD with topology (a), depicted on the right side of Fig. \ref{aprop}. The first one is subleading in the large $N$ limit, whereas each one of the other two gives an expression which is two times a single $\mathcal{N}=4$ SYM flavour flow diagram. This happens because the fundamental loop produces a $N_f= 2\, N$ factor. Taking the difference between diagrams with topology (a) in $\mathcal{N}=4$ SYM and in $\mathcal{N}=2$ SCQCD we thus obtain two times a single flavour flow diagram of $\mathcal{N}=4$ SYM. \vspace{0.2cm}
\begin{figure} [ht]
\hspace{3.6cm}
 \begin{minipage}{8cm}
\includegraphics[scale=0.46]{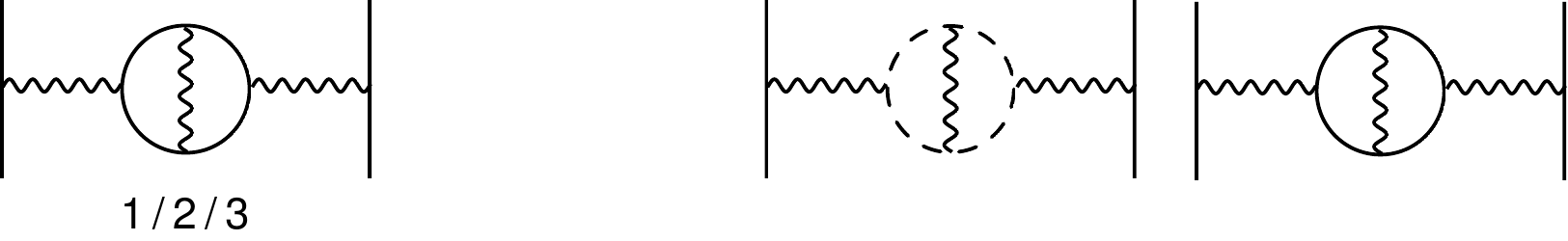}
\end{minipage}  
\caption{$\mathcal{N}=4$ SYM diagrams of topology (b) on the left side and $\mathcal{N}=2$ SCQCD ones on the right side. In $\mathcal{N}=4$ SYM we have three flavour flows while in $\mathcal{N}=2$ SCQCD only the diagram with the adjoint loop survives the large $N$ limit.}  \label{bprop}
\end{figure}

The analysis of the other diagrams is straightforward; as an example we picture in Fig. \ref{bprop} the case of diagram (b) of Fig. \ref{proptot}. Eventually it is found that for every diagram of Fig. \ref{proptot} the difference between their evaluation in $\mc N=4$ SYM and $\mc N=2$ SCQCD is always twice a single flavour flow diagram of $\mathcal{N}=4$ SYM.  We should also consider an additional color factor 2 for diagrams (a), (b) and (c) and a color factor 4 coming from the contractions giving diagram (d) and its permutations. After performing D-algebra, projecting to components and solving the bosonic Feynman integrals we find that 
\footnote{We systematically omit an overall $g^6N^2$ in all intermediate steps of the calculation.}
\vspace{0.2cm}
\begin{align}
(a) & = \frac{4t}{s^{1+2\epsilon}} \bigg[ \frac{1}{2} \, G[1,1]^2 -G[1,1]G[1,\epsilon]  \bigg]  \label{eq:aprop}   \\
(b) & = \frac{4t}{s^{1+2\epsilon}}  \bigg[-\frac{1}{2} \, G[1,1]^2 + 2 G[1,1]G[1,1+\epsilon]   \bigg]    -2 t \,\, \begin{minipage}{50px} \includegraphics[scale=0.09]{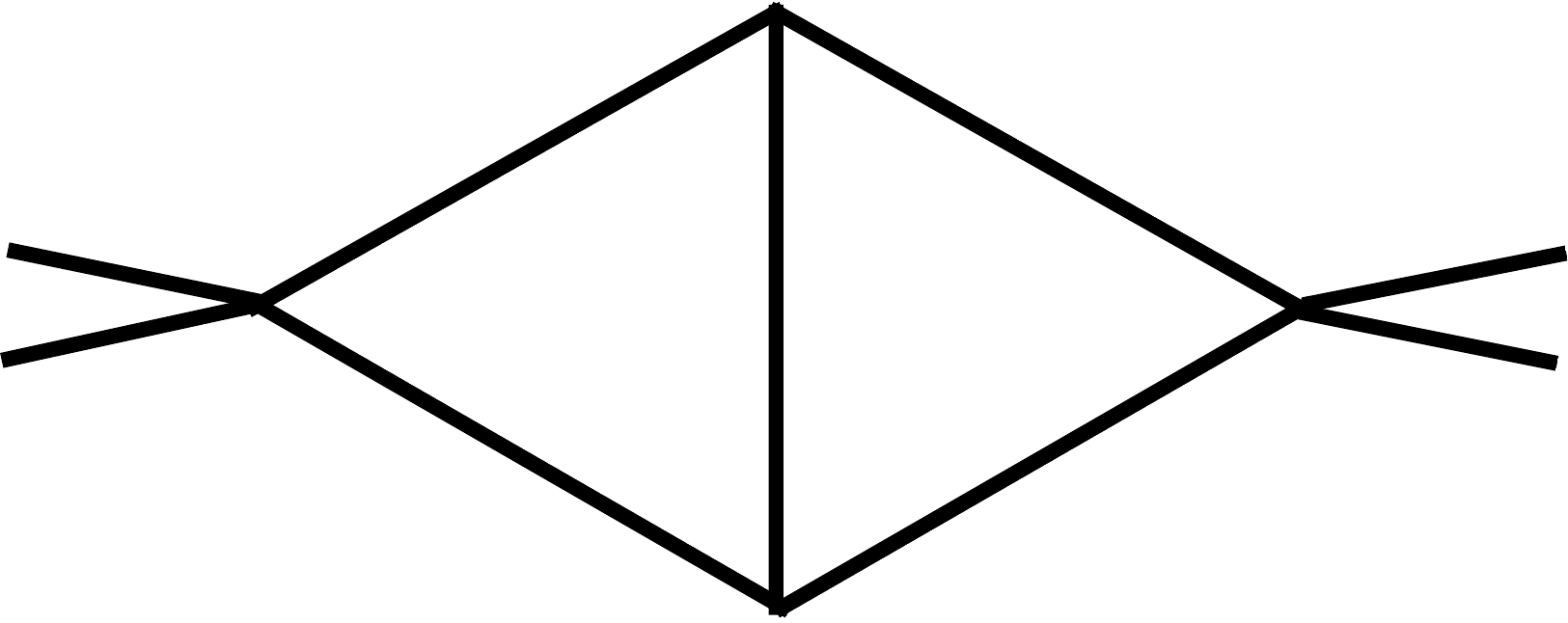} \label{cprop}
\end{minipage}    \\
(c) & = -\frac{4t}{s^{1+2\epsilon}} \,\, G[1,1]G[1, \epsilon]     \\
(d) & = \frac{4t}{s^{1+2\epsilon}}  \bigg[ 2 \, G[1,1] G[1,\epsilon] - 2 G[1,1]G[1,1+\epsilon]   \bigg]    \label{eq:dprop}
\end{align}
where the G--functions are the standard one--loop bubble integrals (see e.g. \cite{Leoni:2014fja} for the explicit definition). We have left the pictorial representation of the double triangle integral in (\ref{cprop}) instead of putting its expansion in terms of G-functions because it is immediate to see that it is the only term surviving to the sum of all the diagrams \eqref{eq:aprop}--\eqref{eq:dprop}. Moreover the double triangle integral is at a glance known to be finite in four dimensions and proportional to the Riemann $\zeta(3)$. We thus obtain the following overall difference between the diagrams containing the two--loop propagator correction in the two models \vspace{0.2cm}
\begin{equation}
\begin{minipage}{1cm}
\includegraphics[scale=0.2]{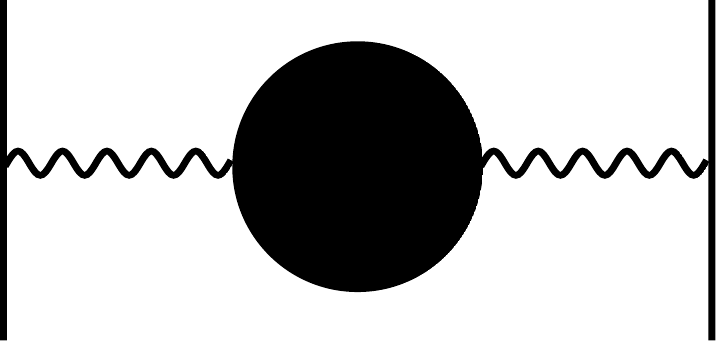}
\end{minipage}\hspace{0.7cm} =\,\, -2 t \,\, \begin{minipage}{1cm}
\includegraphics[scale=0.09]{Dtria6.pdf}
\end{minipage}  \hspace{0.8cm}  =  -\frac{12 t}{s^{1+2\e}} \frac{e^{-2 \e \g_E }}{(4\p)^{4-2\e}} \,\,\zeta (3) + \mathcal{O}(\epsilon) \label{propcont}
\end{equation}
Notice that this contribution to the amplitude difference does not exhibit maximum degree of transcendentality, which should be 4 for the $\mathcal{O}(\epsilon^0)$ terms at two loops \footnote{Strictly speaking, in number theory the transcendentality of $\zeta(n)$ with odd $n$ has not been proven yet. Here, in analogy with the case of even argument,  we assume a degree of transcendentality $n$ for  the $\zeta(n)$ numbers.}. 

\subsection{Vertex corrections}

\begin{figure} [ht]
\centering
 \includegraphics[scale=0.85]{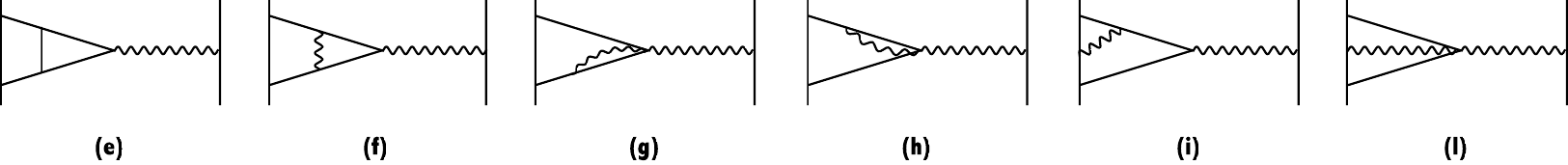}  \vspace{-0.2cm}
\caption{Diagram topologies containing the two--loop vertex corrections which give a non vanishing contribution to the diagrammatic difference.} 
\label{verttot}
\end{figure}

The second class of diagrams we analyze is given by  two--loop vertex corrections of the tree level process. The diagram topologies which might give a non--vanishing contribution to the difference are listed in Fig. \ref{verttot}. 

\begin{figure} [ht]
\hspace{1.8cm}
 \begin{minipage}{4cm}
\includegraphics[scale=0.24]{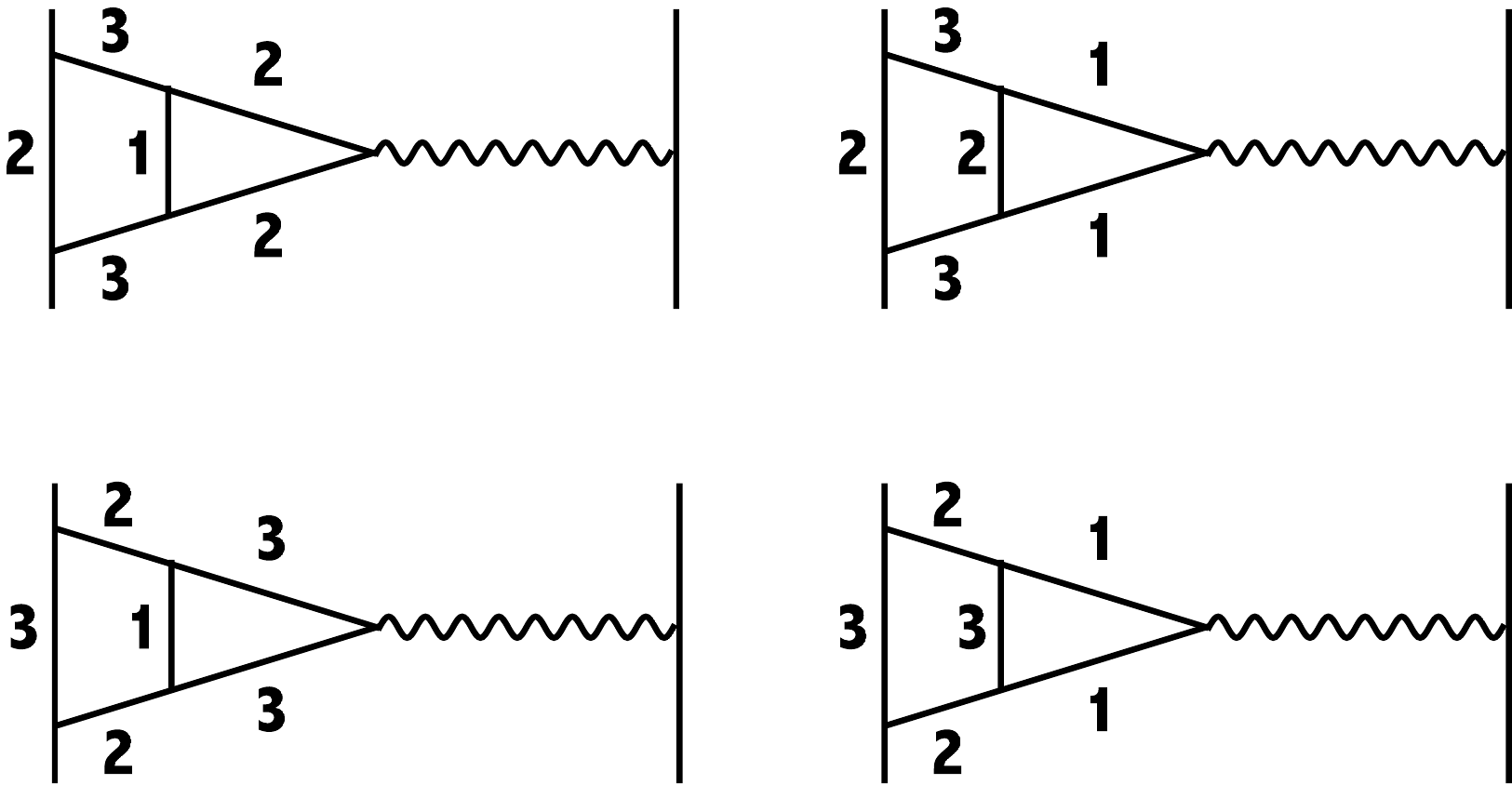}
\end{minipage}  
\hspace{2cm}
\begin{minipage}{4cm}
\includegraphics[scale=0.30]{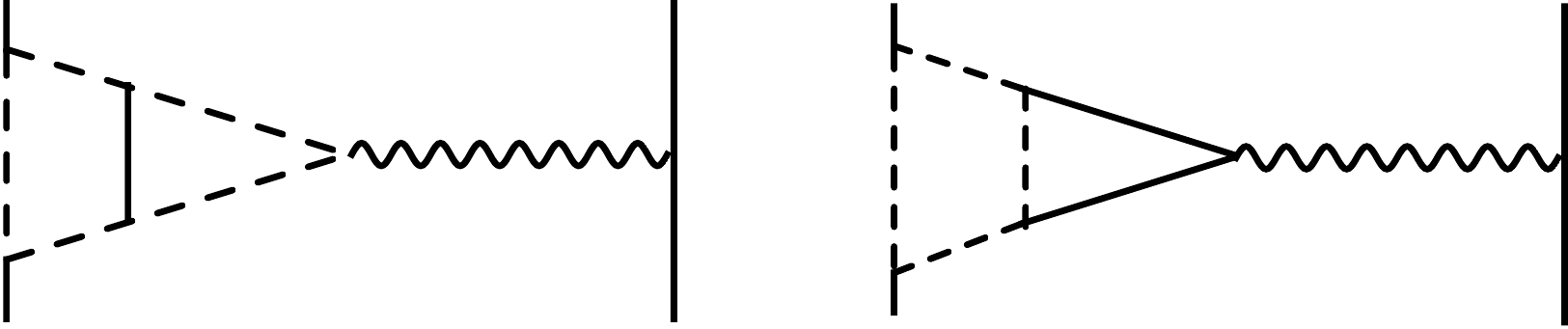}
\end{minipage}
\caption{Flavour flows of $\mathcal{N}=4$ SYM diagrams with topology (e) on the left side and $\mathcal{N}=2$ SCQCD ones on the right side.} \label{evert}
\end{figure}

In $\mathcal{N}=4$ SYM the diagram (e) has four independent flavour flows, as shown in the left side of Fig.  \ref{evert}. On the other hand in $\mathcal{N}=2$ SCQCD we have two diagrams, as shown in the right side of Fig. \ref{evert}. The first one is subleading in the planar limit, while the second one gives two times a single flavour diagram of  $\mathcal{N}=4$ SYM, again because of the presence of a fundamental matter loop. Taking the difference, we are then left with two times a single flavour contribution of $\mathcal{N}=4$ SYM. 

The diagram (f) of Fig. \ref{verttot} has two flavour choices in  $\mathcal{N}=4$ SYM while in  $\mathcal{N}=2$ SCQCD  we only have the subleading diagram. A similar reasoning can be applied to diagrams (g) and (h). Therefore for all the diagrams (e), (f), (g) and (h) we obtain a contribution to the amplitude difference given by twice a single flavour diagram of $\mathcal{N}=4$ SYM. After the first steps of D-algebra diagrams (i) and (l) can be quickly shown to vanish on--shell. Taking into account an additional factor 2 by considering also the correction to the right vertex of the tree--level diagram we obtain after D-algebra and projections 
\begin{align}
(e) &=   \begin{minipage}{45px} \includegraphics[width=1.3cm]{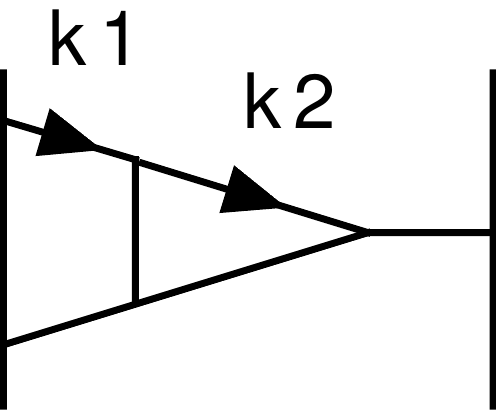} \vspace{0.15cm}  \end{minipage}\, 4 \,\, \bigg(  - k_1^2 \, \Tr(k_2 p_2 p_1 p_4) + k_2^2\, \Tr(k_1 p_3 p_4 p_1)  - s\,  \Tr(k_1 k_2 p_4 p_1) \bigg) \non \\
(f) &=  \begin{minipage}{45px} \includegraphics[width=1.3cm]{TraceVert.pdf} \vspace{0.15cm} \end{minipage}\, 4 \,\, \bigg(   k_1^2 \, \Tr(k_2 p_3 p_4 p_1) + (k_2-p_3-p_4)^2 \, \Tr(k_1 p_3 p_4 p_1) \,\,  +  \non \\ &  \hspace{3cm}  - s  \, \Tr(k_1 p_3 p_4 p_1)  + s\, \Tr(k_1 k_2 p_4 p_1) \bigg)\non \\
(g) &=   \begin{minipage}{45px} \includegraphics[width=1.3cm]{TraceVert.pdf} \vspace{0.15cm}  \end{minipage}\, 4 \,\, \bigg(t\,  k_2^2 (k_1 + p_1+p_2)^2 - k_2^2 \Tr((k_1+p_1+p_2 )p_2 p_1 p_4)  \bigg) \non \\
(h) &=  \begin{minipage}{45px} \includegraphics[width=1.3cm]{TraceVert.pdf} \vspace{0.15cm} \end{minipage}\, 4 \,\,  \bigg(t\, k_1^2 (k_2 - p_3-p_4)^2 - (k_2 - p_3-p_4)^2\Tr(k_1 p_3 p_4 p_1)  \bigg)\non 
\end{align}
We found convenient to combine the trace structure of diagrams (e) with (f) and (g) with (h). Completing the squares we can cast their sum in terms of the following combination of scalar integrals 
\begin{align} 
(e)+(f) & = 4 \,\, \bigg( -\frac{t}{s} \,\begin{minipage}{50px} \includegraphics[width=1.5cm]{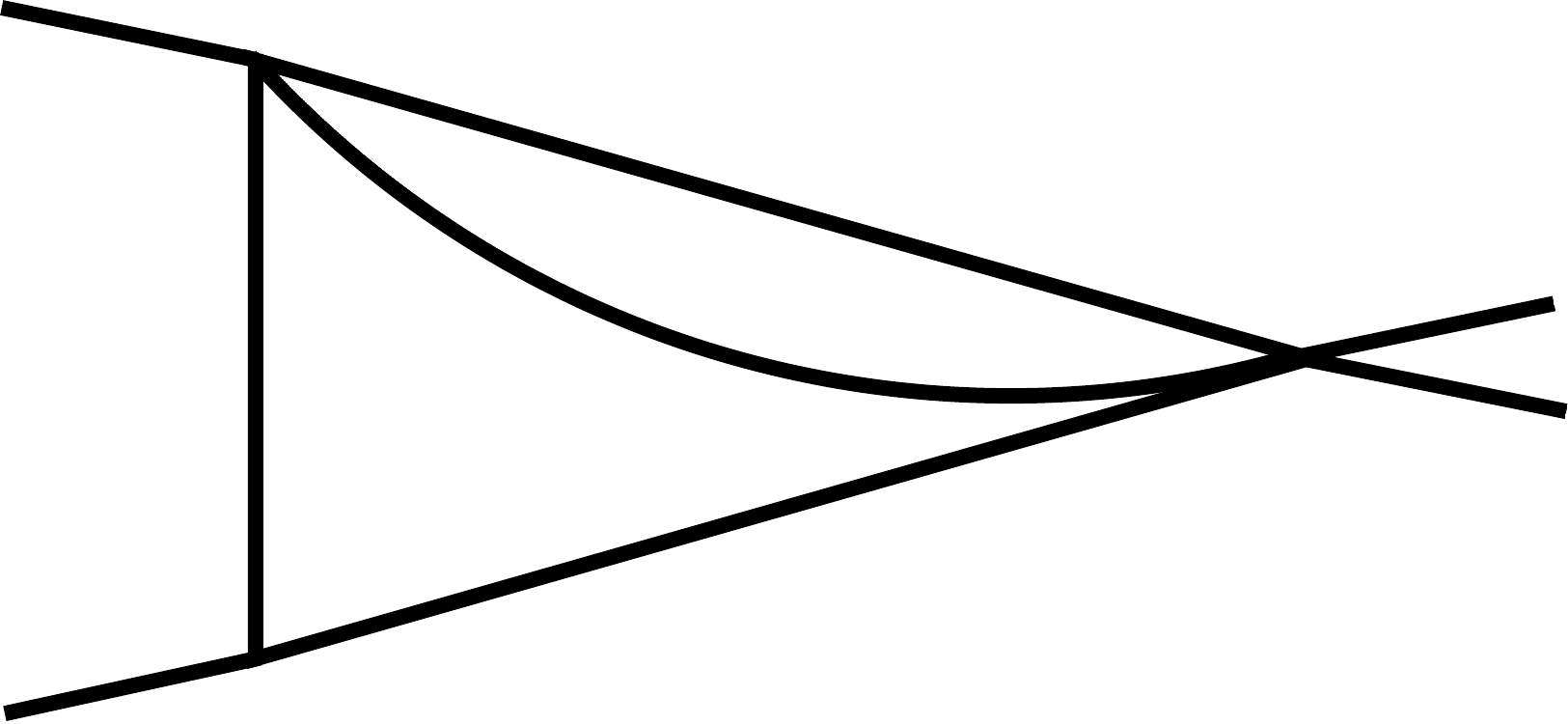} \end{minipage}\, +\,\frac{s+ 2 t}{2}  \,\,\,\,\begin{minipage}{50px} \includegraphics[width=1.5cm]{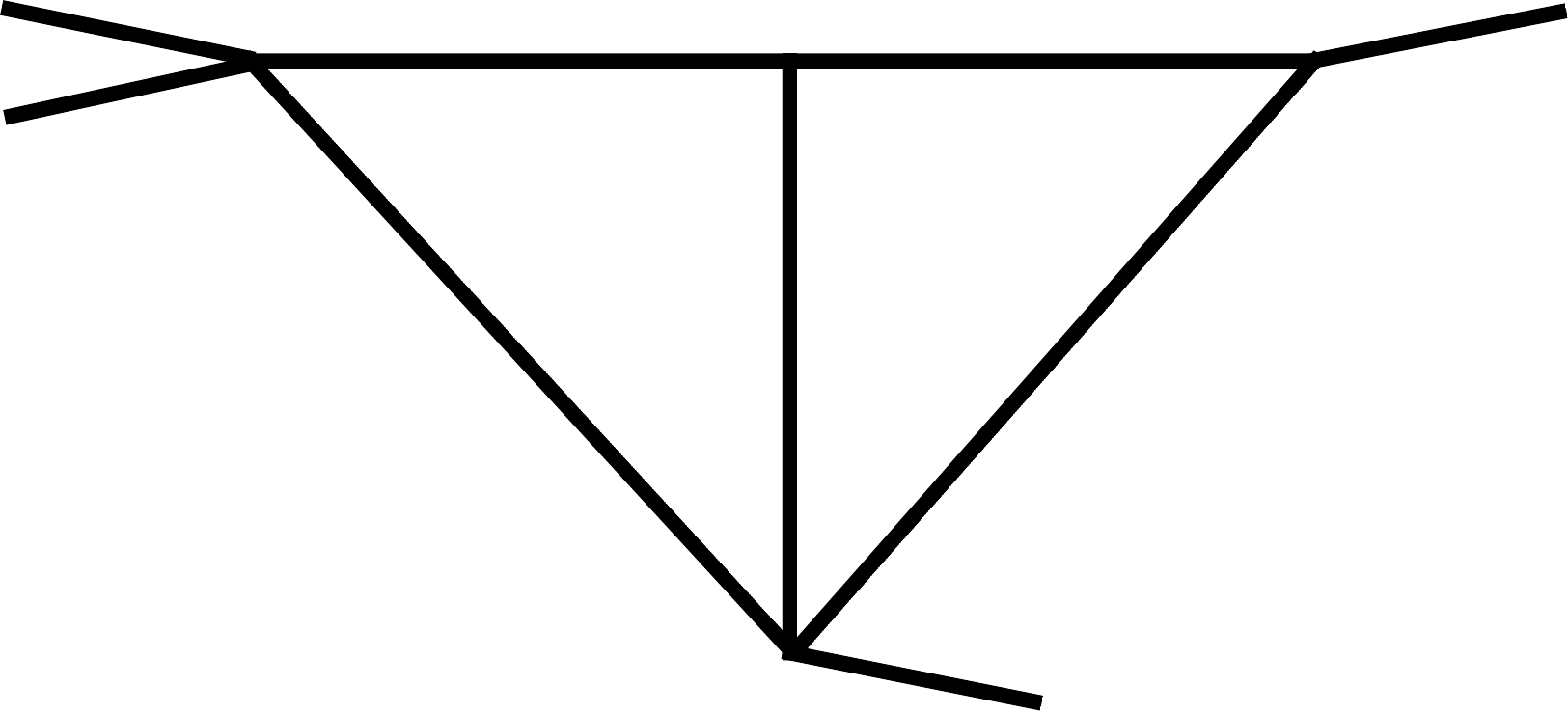} \end{minipage} +  \,\,\,\begin{minipage}{50px} \includegraphics[width=1.6cm]{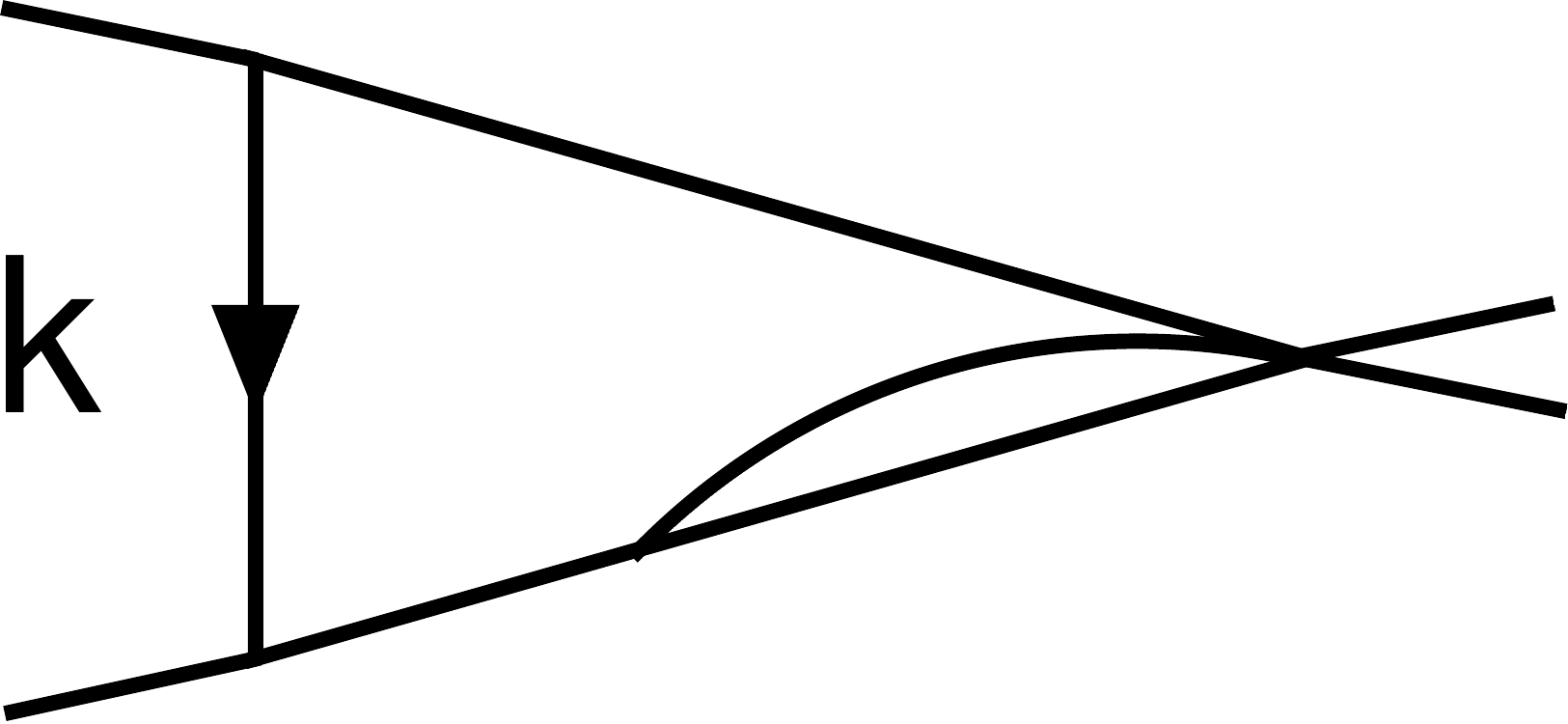} \end{minipage}(k-p_4)^2 \,\, + \non \\[0.2cm] & \hspace{1cm} -\frac{s}{2}  \,\,\,\begin{minipage}{50px} \includegraphics[width=1.5cm]{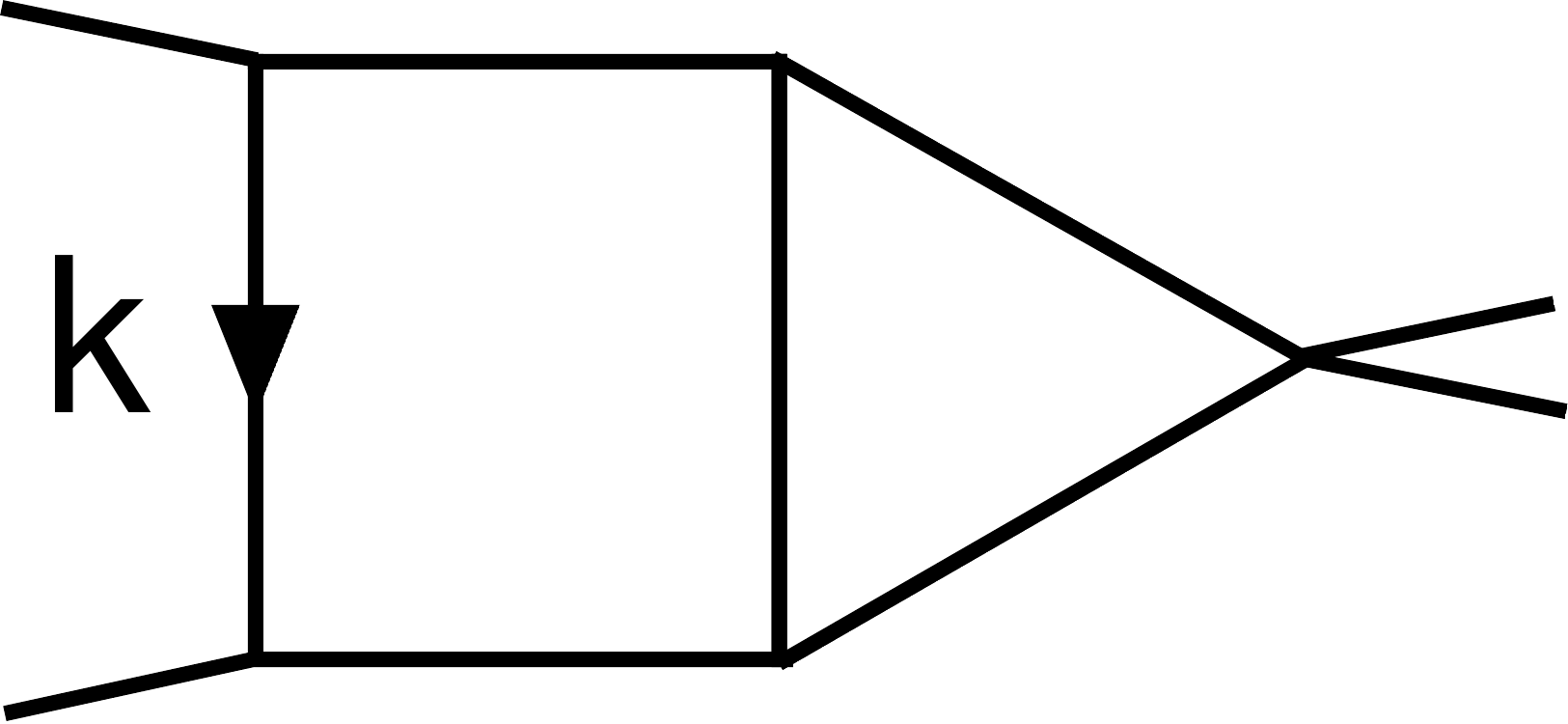} \end{minipage}(k-p_4)^2\bigg)    \\[0.25cm] 
(g)+(h) & = 4 \,\, \bigg( \frac{t}{s} \,\begin{minipage}{50px} \includegraphics[width=1.5cm]{TNew.pdf} \end{minipage}-  \,\,\,\begin{minipage}{50px} \includegraphics[width=1.6cm]{MNew.pdf} \end{minipage} (k-p_4)^2\bigg) 
\end{align}
Taking the sum, the remaining contribution can be expanded in terms of master integrals using expansions  which have been obtained through the {\it Mathematica} package  {\bf FIRE} \cite{Smirnov:2008iw}. The final result turns out again to be proportional to the double triangle integral
\begin{equation}
\begin{minipage}{2.4cm}
\includegraphics[scale=0.23]{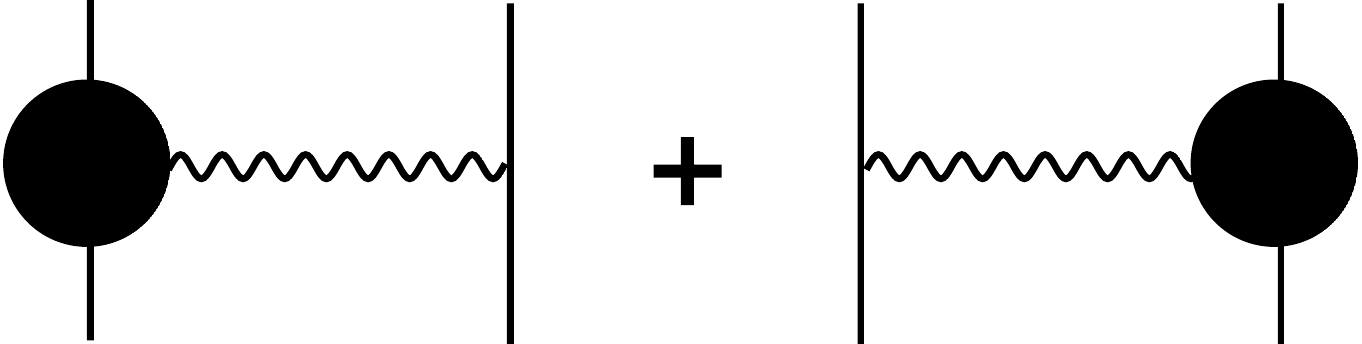}
\end{minipage}\hspace{0.8cm} =\,\, 4 t \,\, \begin{minipage}{1cm}
\includegraphics[scale=0.09]{Dtria6.pdf}
\end{minipage}  \hspace{0.8cm}  = \,\,  \frac{24 t}{s^{1+2\e}} \frac{e^{-2 \e \g_E }}{(4\p)^{4-2\e}} \,\,\zeta (3) + \mathcal{O}(\epsilon) \label{vertcont}
\end{equation}
Again, this contribution to the difference of amplitudes does not exhibit maximum degree of transcendentality. From equations  (\ref{propcont}) and  (\ref{vertcont}) we also notice that the propagator  and vertex  dressings of the tree level process only give constant contributions to the amplitude difference. This kind of terms does not affect the potential presence of the duality with Wilson loops.

\subsection{Genuine two--loop diagram}\label{dboxsec}

\begin{figure} [ht]
\centering
 \includegraphics[scale=0.27]{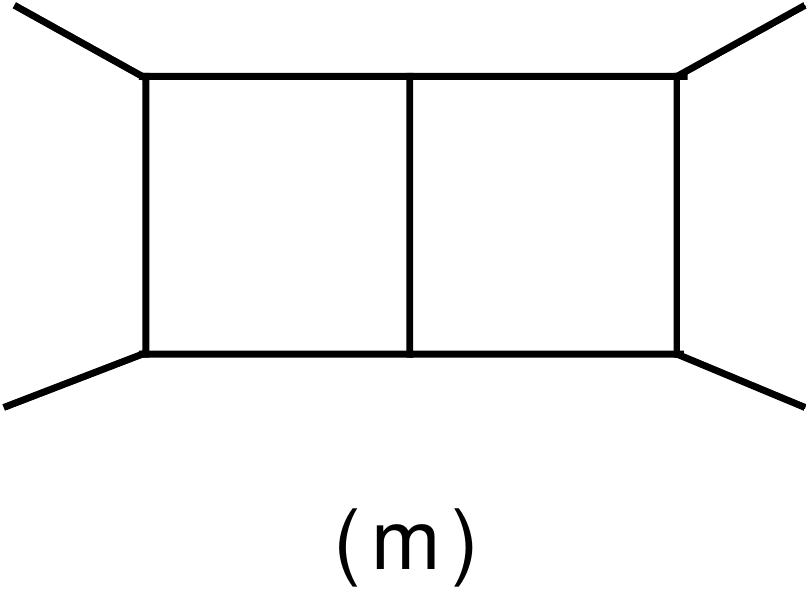} \vspace{-0.3cm}
\caption{Two-loop ladder topology contributing to the diagrammatic difference.} \label{dbox}
\end{figure}

There is a third class of diagrams which have to be considered, which are the genuine two--loop diagrams. In Fig. \ref{dbox} we depicted the ladder diagram, which is the only diagram topology which contributes to the amplitude difference.
\begin{figure} [ht] \hspace{2.5cm}
 \begin{minipage}{4.5cm}
\includegraphics[scale=0.26]{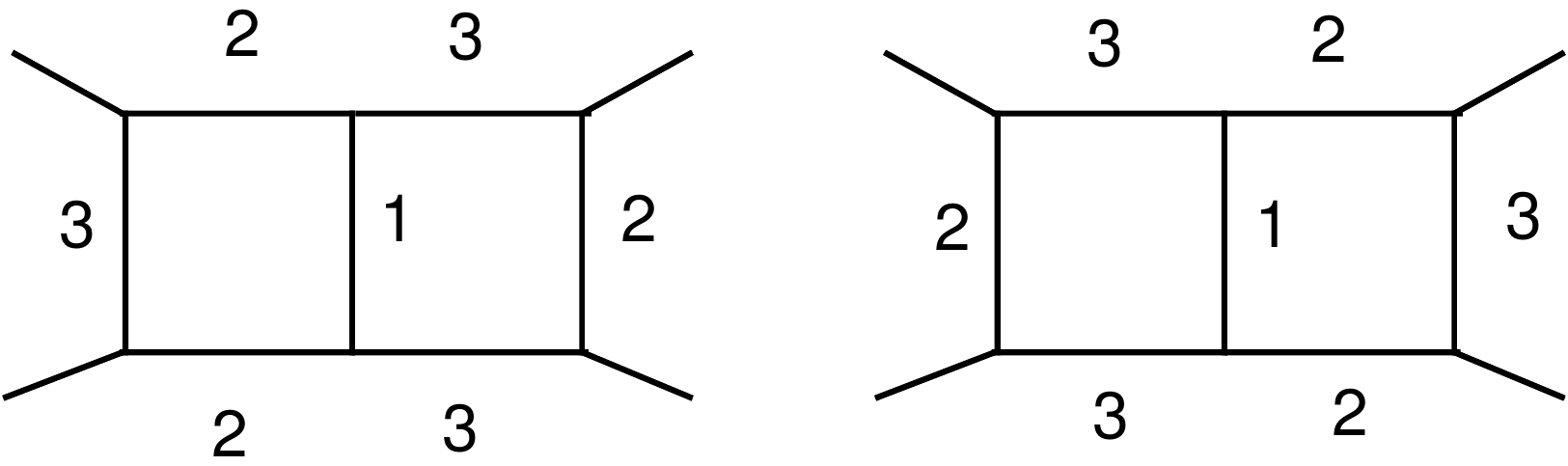}
\end{minipage}  
\hspace{2cm}
\begin{minipage}{4.5cm} 
\includegraphics[scale=0.24]{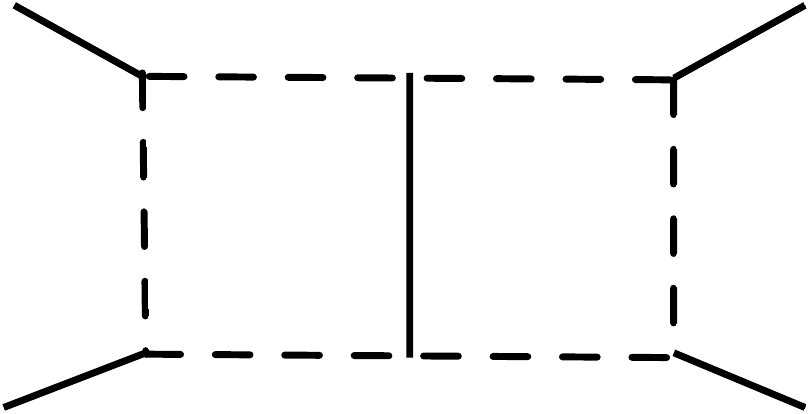}
\end{minipage} 
\caption{Flavour flows of $\mathcal{N}=4$ SYM diagrams with topology (m) on the left side and the $\mathcal{N}=2$ SCQCD one on the right side.} \label{dbox2}
\end{figure}\\
In $\mathcal{N}=4$ SYM there are two flavour flows for the diagram (m), as shown in the left side of Fig. \ref{dbox2}, whereas in $\mathcal{N}=2$ SCQCD this topology only admits a subleading realization, as shown in the right side of Fig. \ref{dbox2}. After performing D-algebra and projections, we are left with the following combination of bosonic integrals with $\g$-trace numerators 
\begin{align}
(m) & =  \begin{minipage}{50px} \includegraphics[width=1.5cm]{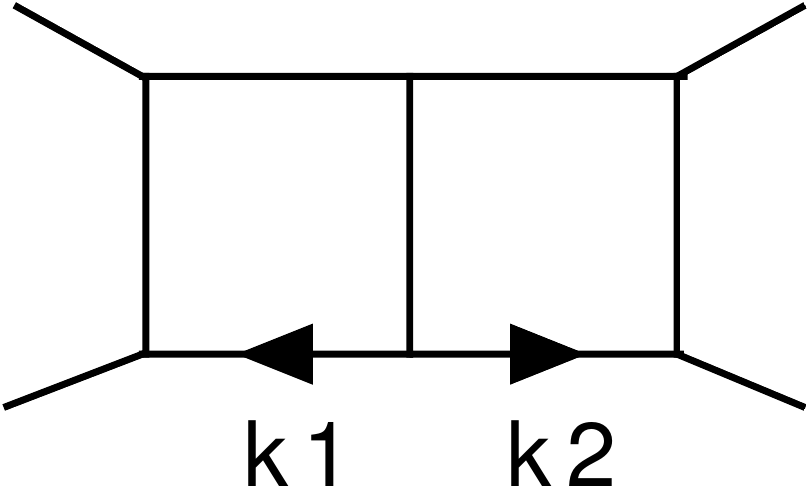} \end{minipage}\,\, 2 \, \, \bigg(- t \,  k_2^2 (k_1-p_1-p_2)^2 + k_2^2\, \Tr\big((k_1-p_1-p_2) p_3 p_4 p_1\big) \,\, + \non \\ & \hspace{4cm} - \Tr\big(k_1 k_2 p_3 p_4 p_1 (k_1-p_1-p_2)\big) \bigg) 
\end{align}
After some algebra and completing the squares we end up with a linear combination of scalar integrals \vspace{0.1cm}
\begin{align} 
(m) = &\, s \begin{minipage}{50px} \includegraphics[width=1.4cm]{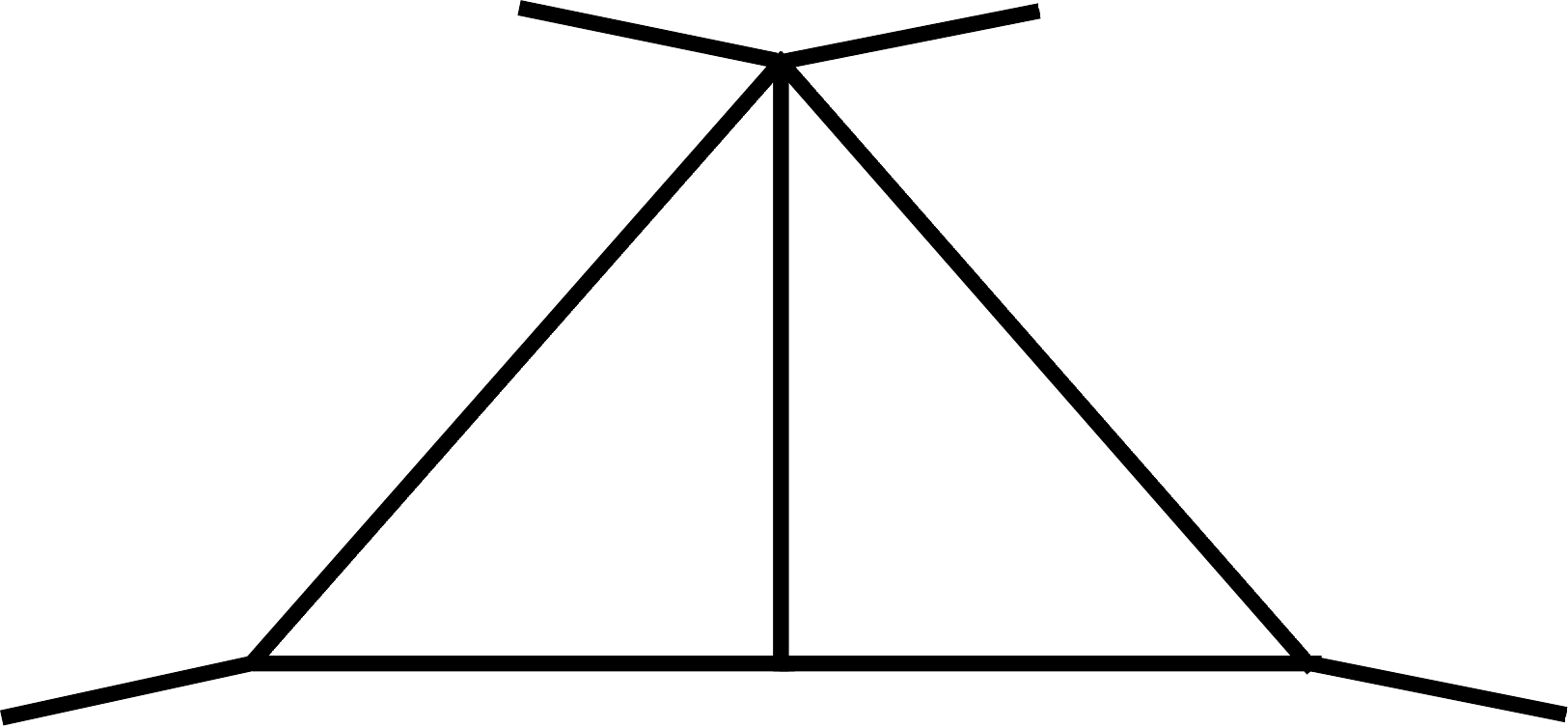} \end{minipage}\!\!\!\!\!\!-2 s \,\,\begin{minipage}{50px} \includegraphics[width=1.4cm]{Dtria4.pdf} \end{minipage} \!\!\! \!\!+2s \,\, \begin{minipage}{50px} \includegraphics[width=1.4cm]{Horhousevec.pdf} \end{minipage} \!\!\!(k-p_4)^2 +2s \,\, \begin{minipage}{50px} \includegraphics[width=1.4cm]{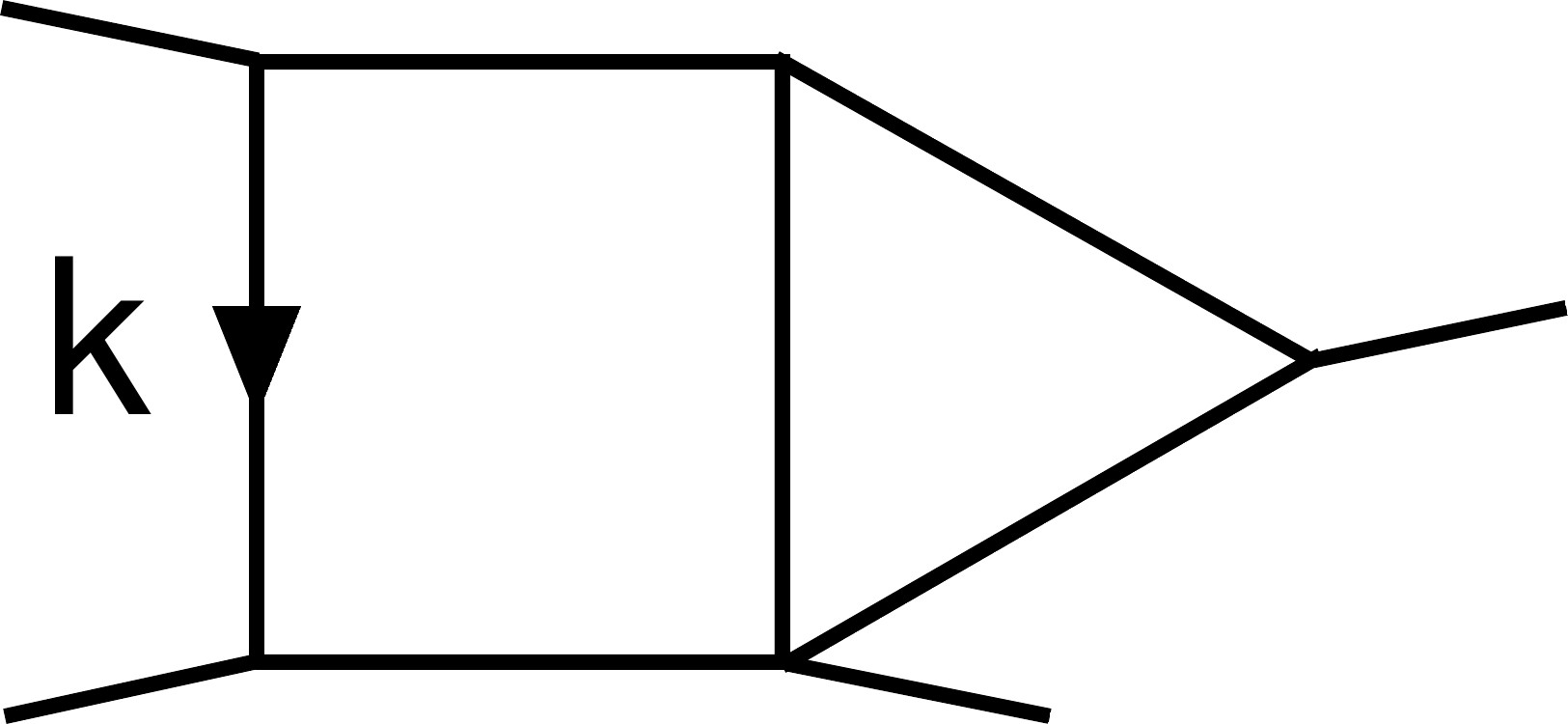} \end{minipage} \!\!\! (k-p_4)^2 \, + \non \\[0.25cm] 
& \hspace{-1,2cm}+ st\,\, \begin{minipage}{1cm}
\includegraphics[scale=0.09]{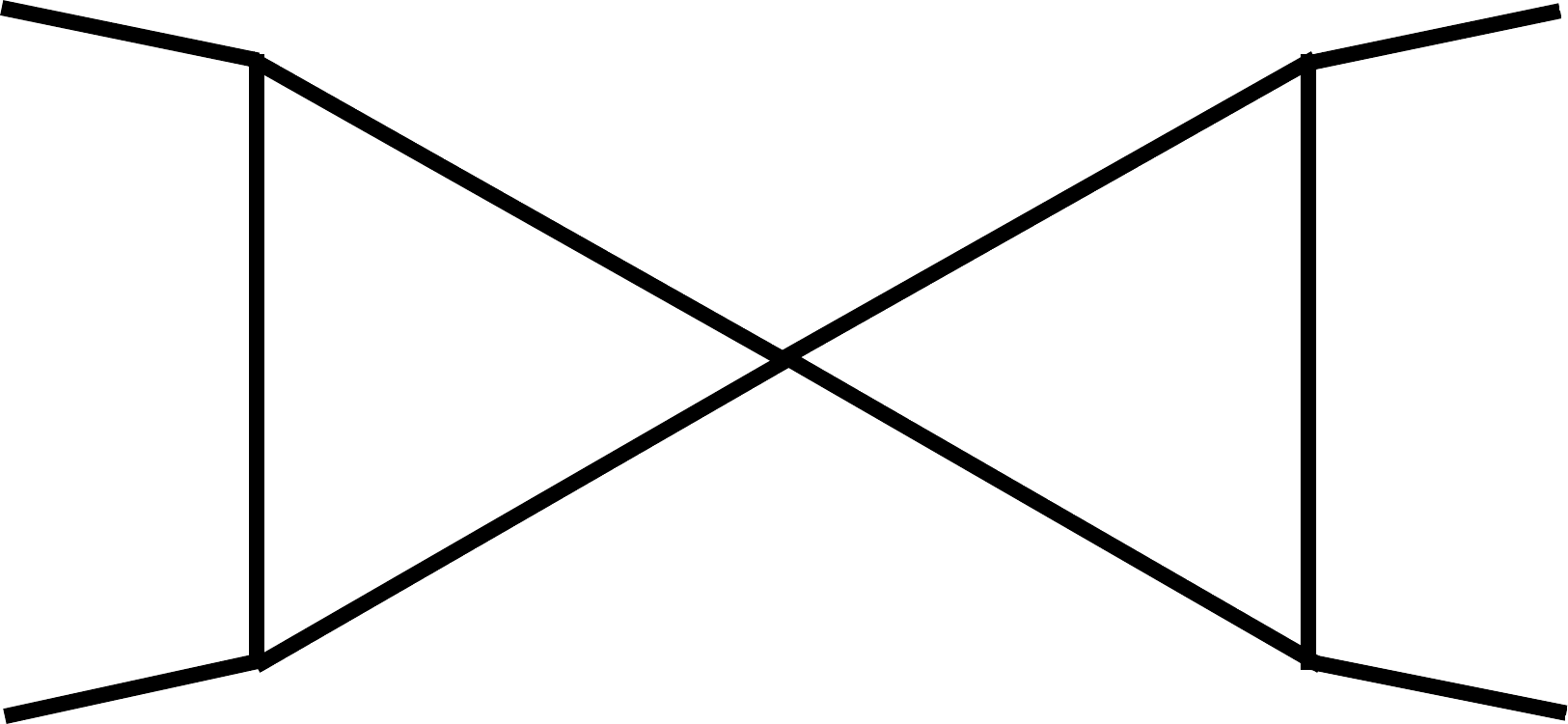}
\end{minipage} \,\,\,\,\,\, -2(s+t) \, \begin{minipage}{50px} \includegraphics[width=1.2cm]{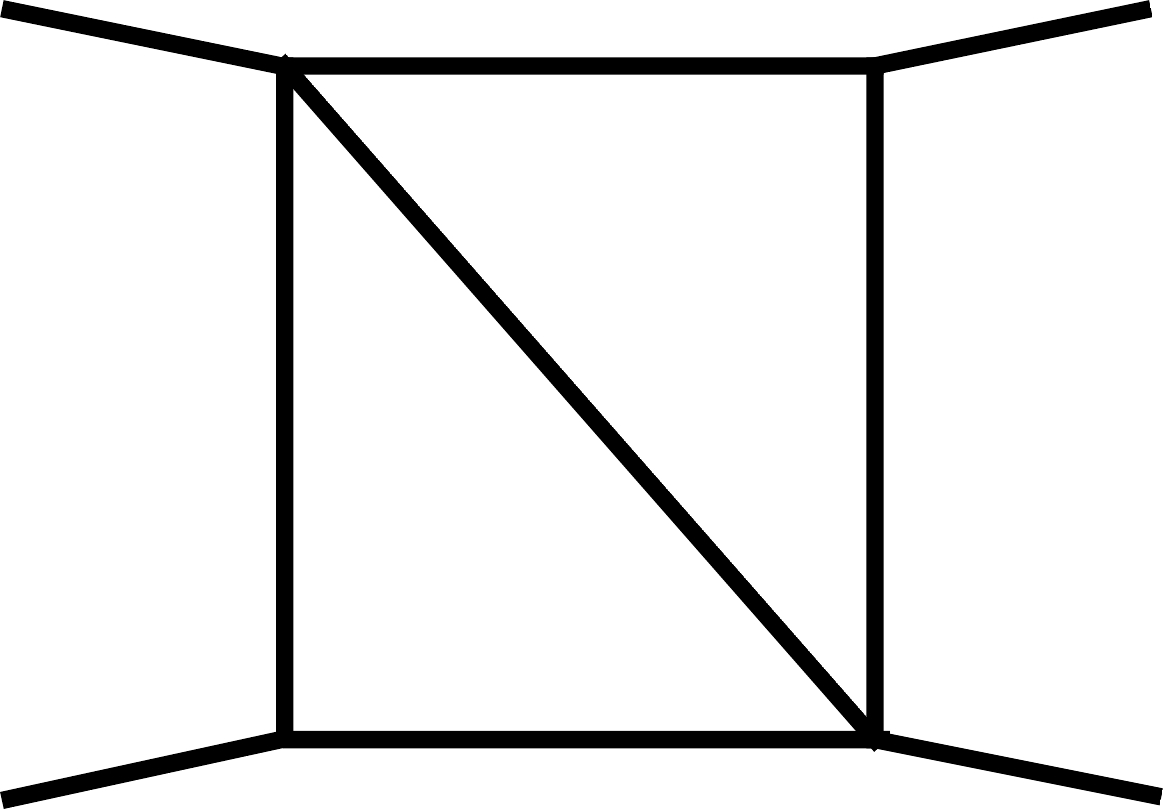} \end{minipage} \hspace{-0.5cm}   -2 s t\begin{minipage}{50px} \vspace{-0.1cm}\includegraphics[width=1.7cm]{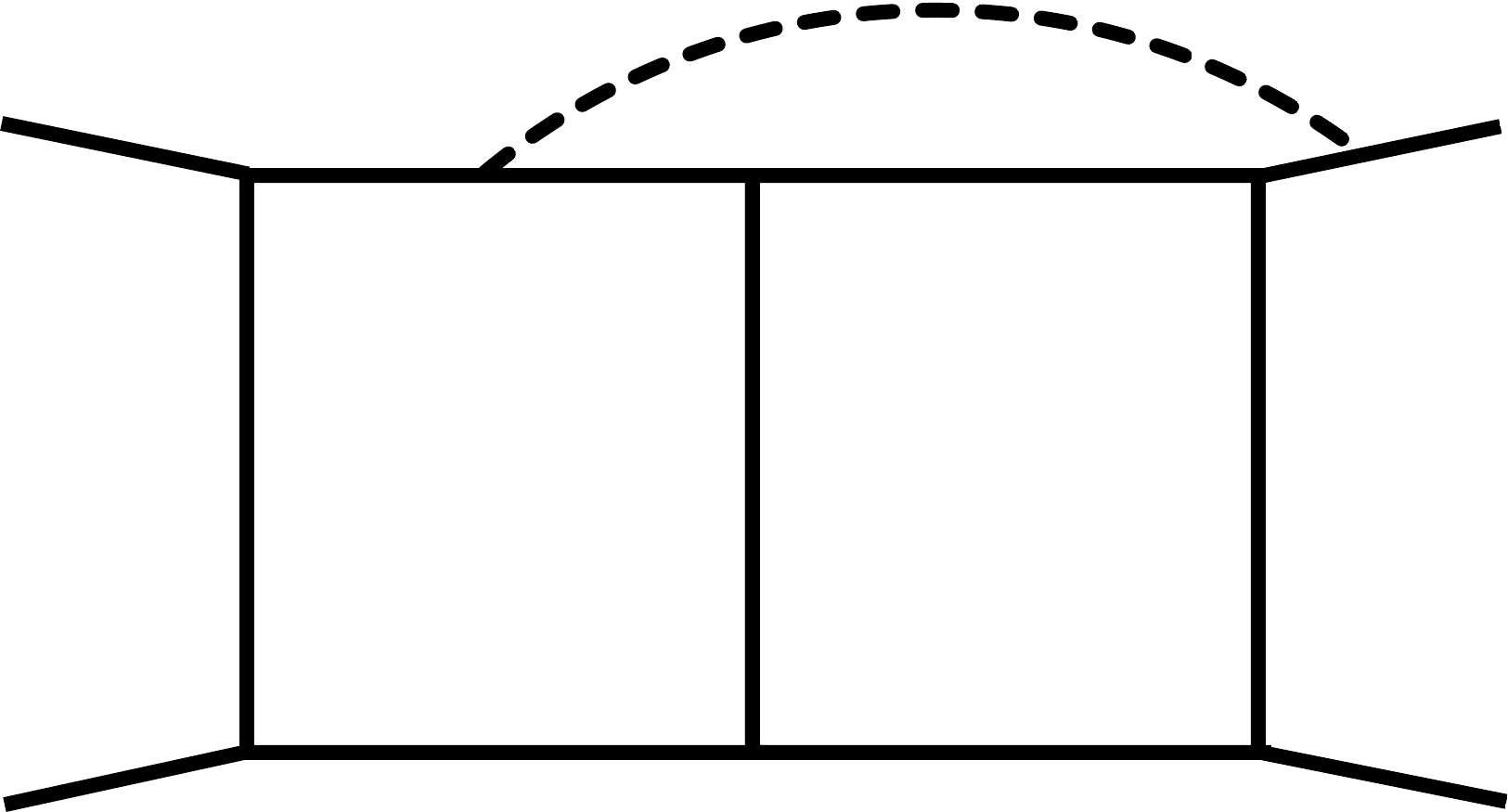} \end{minipage} + s t^2 \,\,  \begin{minipage}{50px} \includegraphics[width=1.4cm]{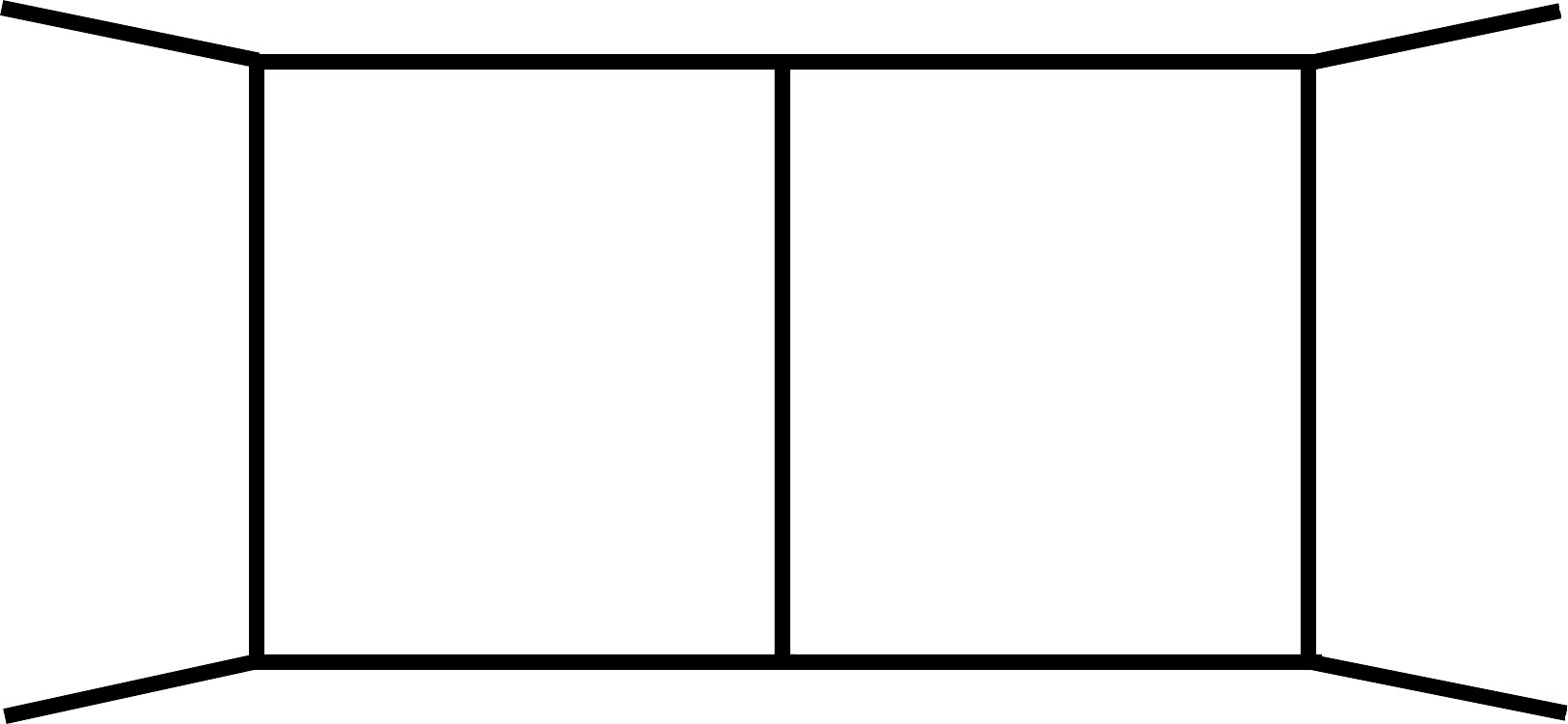}\end{minipage}\!\!\!\! - s \,\,\, \begin{minipage}{50px}\vspace{0.2cm} \includegraphics[width=1.4cm]{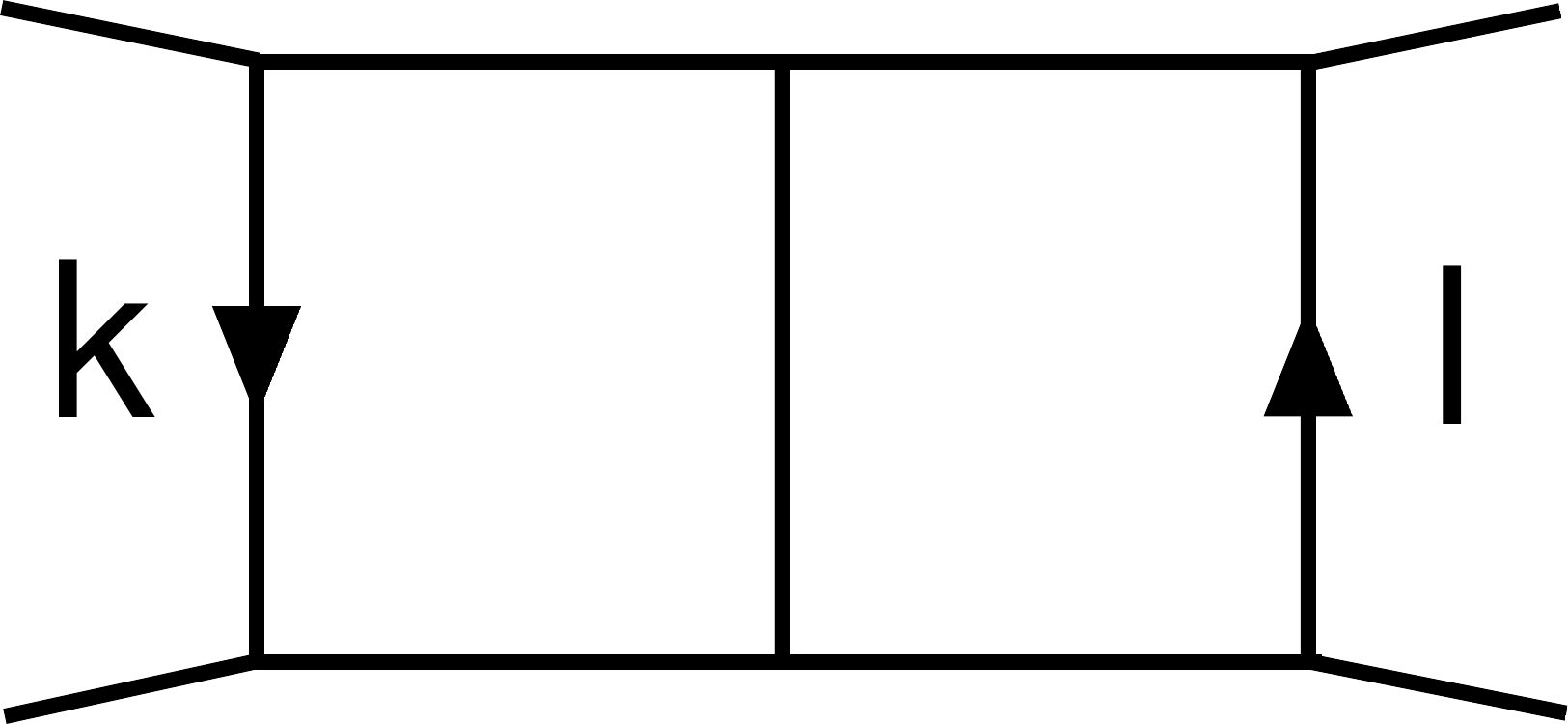}\vspace{0.2cm}\end{minipage} \!\!\!(k-p_4)^2 (l+p_1)^2 \label{boxscalar}
\end{align}
The scalar integrals in (\ref{boxscalar}) can be expanded on the basis of two--loop master integrals (for details see Appendix D.3 of \cite{Leoni:2014fja}). We  finally obtain  \vspace{0.2cm}
\begin{align} \label{Dboxmaster}
\mc (m)  = & - \frac{1}{2}\, s^2 t \,\, \begin{minipage}{50px} \includegraphics[width=1.7cm]{Ilad.pdf} \end{minipage}  + \frac{3}{2} \, s^2  \,\begin{minipage}{50px} \vspace{-0.1cm}\includegraphics[width=1.7cm]{Ivlad.pdf} \end{minipage}+ 7\, (s+t) \,\,\begin{minipage}{50px} \includegraphics[width=1.2cm]{Idiag.pdf} \end{minipage} \hspace{-0.5cm}- 8\, a s \,\, \begin{minipage}{50px} \includegraphics[width=1.3cm]{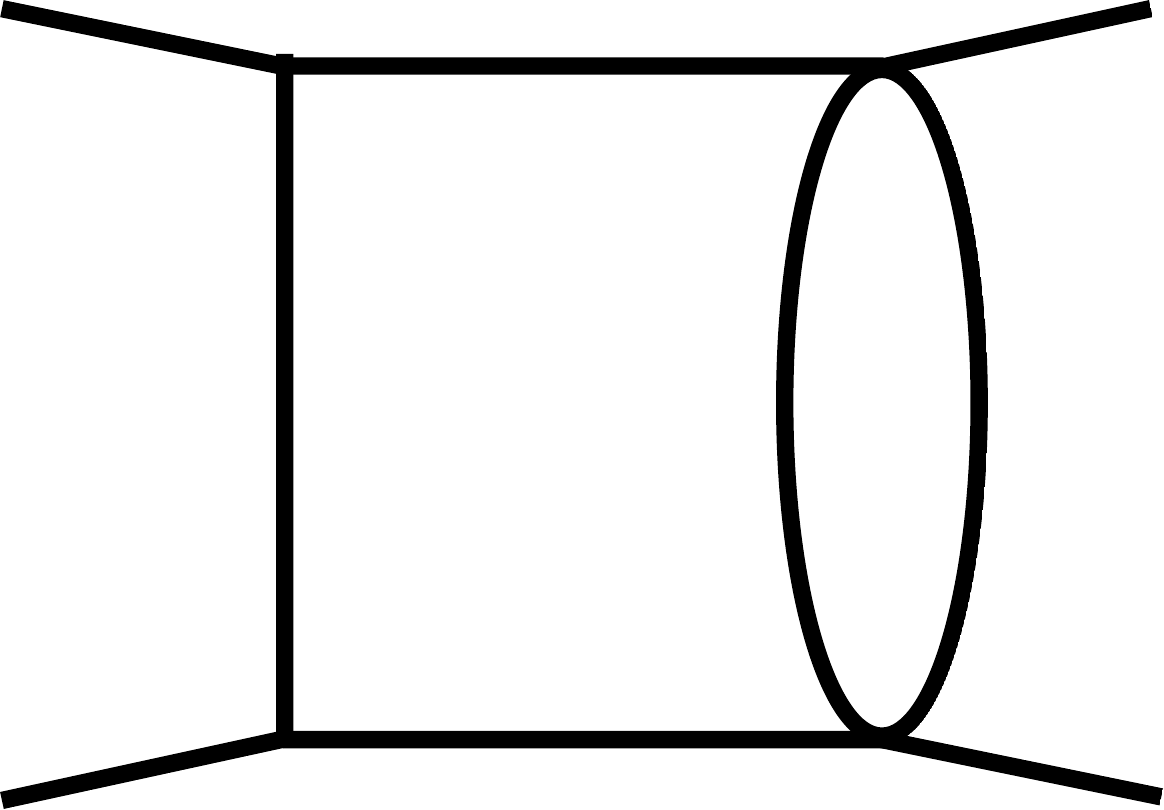} \end{minipage}  \hspace{-0.5cm}\,\,\, \,\,+ \non \\[0.2cm] &  -6\, a^2 \,\, \begin{minipage}{50px} \includegraphics[width=1.6cm]{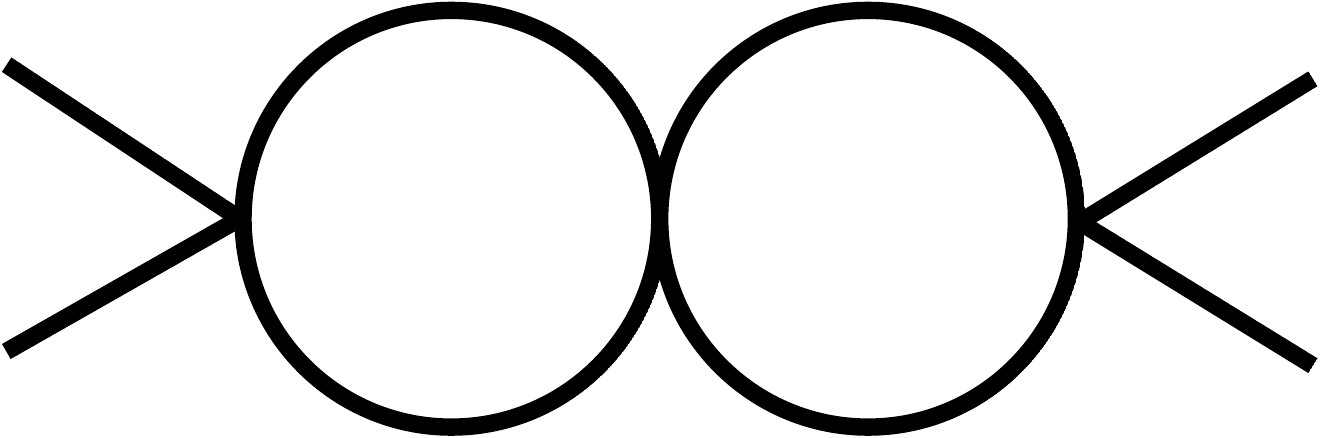} \end{minipage} \,   + \frac{2 c }{s} \,\,\, \begin{minipage}{50px} \includegraphics[width=1.4cm]{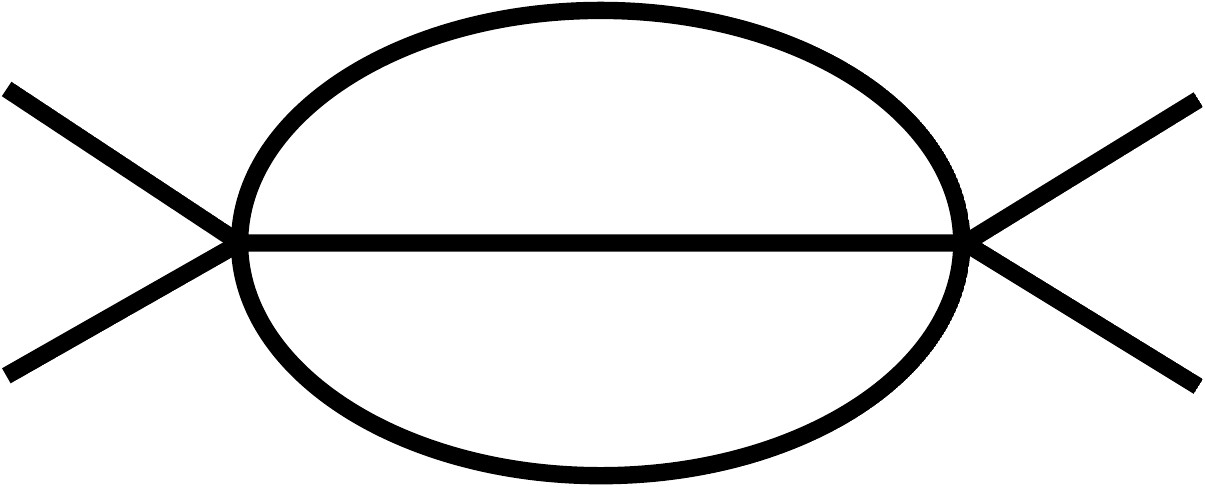} \end{minipage}\!\!- \frac{9 c}{t} \,\,\, \begin{minipage}{50px} \includegraphics[angle=90,width=0.5cm]{Isunset.pdf} \end{minipage} \hspace{-1cm}-\frac{17 b}{2}  \,\,\, \begin{minipage}{50px} \includegraphics[width=1.4cm]{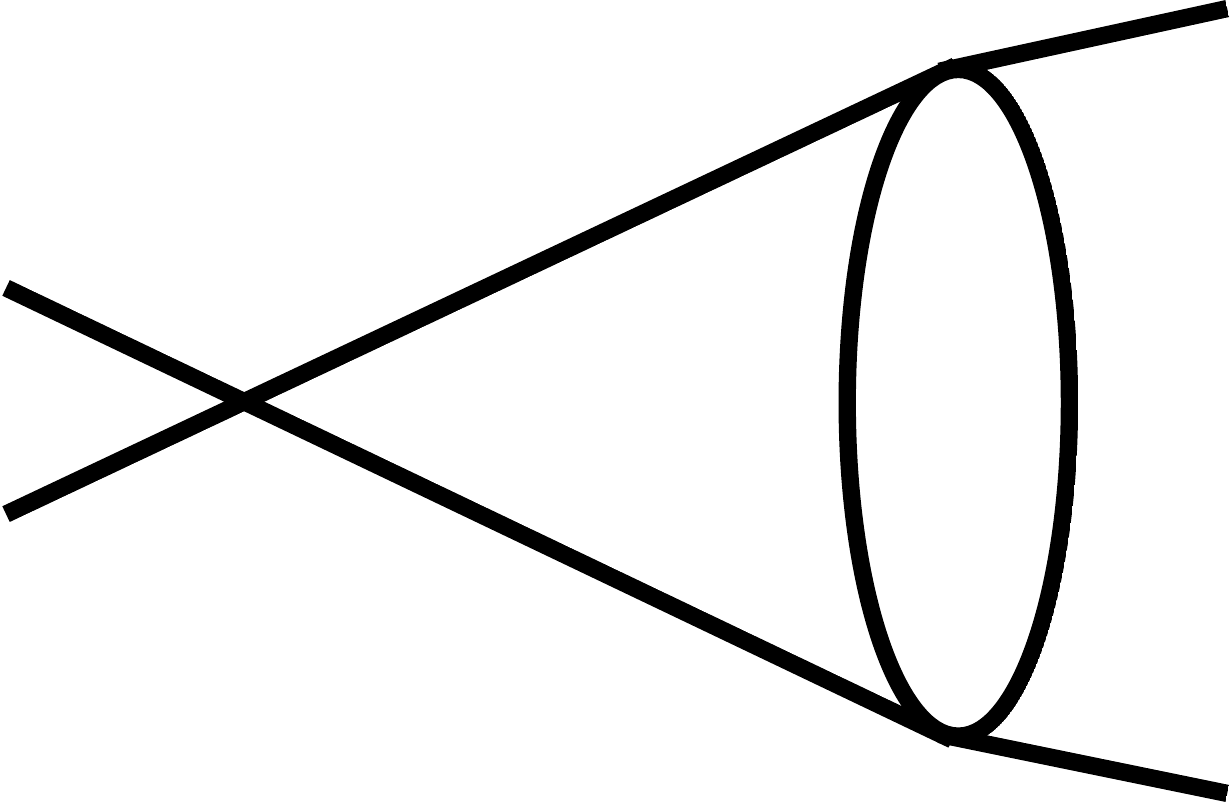} \end{minipage}  
\end{align}
where we defined the coefficients \,
$
a  = - \frac{1-2 \e}{2\e},
\,\, b  =  \frac{(1-2\e)(1-3\e)}{2\e^2},
\,\, c  = -\frac{(1-2\e)(1-3\e)(2-3\e)}{2\e^3}.
$
Expanding the master integrals of eq. \eqref{Dboxmaster} in terms of the dimensional regularization parameter $\e$ gives
\vspace{0.1cm}
\begin{align}
\!(m) &= \frac{2}{s^{2\e}} \frac{e^{-2 \e \g_E }}{(4\p)^{4-2\e}} \bigg[ \frac{1}{6} \ln (1+x) \big(2 \pi^2 \ln x -3 (\pi^2+\ln^2 x) \ln (1+x) \big)+\frac{ \pi^2}{3}\Li{2}{-x}  + \non \\
& +2 S_{2,2}(-x)  - 2  \ln x \, S_{1,2}(-x)-2\ln(1+x)\big(\ln x\, \Li{2}{-x}-\Li{3}{-x}+\zeta(3)\big)\bigg]  \label{boxcont}
\end{align}
where $x=t/s$. 
We thus see that even if the master integrals in eq. (\ref{Dboxmaster}) exhibit individually poles up to $1/\e^4$, the non trivial combination of them produces a contribution which is only finite. Nevertheless, this finite contribution depends non trivially on the kinematics and it is responsible for the breaking of the duality with  Wilson loops.

\subsection{Two--loop result}

Combining eq. \eqref{propcont}, \eqref{vertcont} and \eqref{boxcont}, we can write the full two--loop reduced adjoint amplitude in $\mathcal{N}=2$ SCQCD 
\begin{framed} \vspace{-0.3cm}
\begin{align}
 \mathcal{ M}^{(2)}_{\mathcal{N}=2}  &=  \mc M^{(2)}_{\mathcal{N}=4}  + \frac{\lambda^2}{s^{2\e}} \frac{e^{-2 \e \g_E }}{(4\p)^{-2\e}} \bigg[  \,\, \frac{2}{x} \bigg( \frac{1}{6} \ln (1+x) \big(2 \pi^2 \ln x -3 (\pi^2+\ln^2 x) \ln (1+x) \big) +  \non\\
 & + 2 S_{2,2}(-x)  - 2  \ln x \, S_{1,2}(-x)  + \frac{ \pi^2}{3}\Li{2}{-x}   -2\ln(1+x)\big(\ln x\, \Li{2}{-x}+ \non \\
 & -\Li{3}{-x}+\zeta(3)\big)\bigg) + 12 \zeta (3)   \bigg]  \label{result}
\end{align} \vspace{-0.5cm}
\end{framed}
The result is equal to the $\mathcal{N}=4$ SYM one plus a finite part. Since the $\mc N=4$ SYM amplitude is dual conformal invariant we conclude that the two--loop amplitude in $\mc N=2$ SCQCD does not manifest this symmetry. 
The difference of amplitudes in $\mathcal{N}=4$ and in $\mathcal{N}=2$ consists in a finite  expression, which depends non trivially on the kinematic variables $s$ and $t$. This result does not agree with the null difference between the two--loop expectation value of the four--sided Wilson loop in the two models. So we conclude that the scattering amplitude/Wilson loop duality is broken at two loops in $\mc N=2$ SCQCD. 
Nonetheless the difference of amplitudes between $\mathcal{N}=4$ and $\mathcal{N}=2$, except for the irrelevant constant $12 \zeta(3)$, is suppressed at large $x$ by a factor of $1/x$,  where $x=t/s$.
The limit of large $x$ corresponds to the Regge asymptotics of the amplitude. This means that even though the amplitude/Wilson loop duality is broken at two loops, it gets restored in the Regge limit \footnote{We thank G. Korchemsky for raising our attention on this point.}. This behaviour for the four--point amplitude was first observed in QCD \cite{Korchemskaya:1996je, Drummond:2007aua} and in this respect it makes $\mathcal{N}=2$ SCQCD a much closer relative of QCD rather than of $\mc N=4$ SYM, where the amplitude turns out to be Regge exact. 

The finiteness of the difference is consistent with the exponentiation of the infrared poles of amplitudes in $\mc N=4$ SYM and in $\mathcal{N}=2$ SCQCD: the two--loop poles indeed are fixed by the one--loop amplitude, which is identical in the two models. A non finite difference instead would have spoiled the exponentiation.

The result \eqref{result} does not exhibit uniform transcendentality weight due to the presence of the $\zeta(3)$ term, meaning that even in the adjoint sector of $\mc N=2$ SCQCD the maximum transcendentality principle is violated.

\section{Comments and conclusions} \label{sec4}

We computed the two--loop four--point reduced amplitude in the adjoint sector of $\mc N=2$ SCQCD and the result is presented in eq. \eqref{result}. We found that the difference between the $\mc N=4$ SYM and the $\mc N=2$ SCQCD amplitudes is finite and depends non trivially on the kinematic variables. This implies that the $\mc N=2$ SCQCD amplitude is not dual conformal invariant and the scattering amplitude/Wilson loop duality is broken in $\mc N=2$ SCQCD, since the expectation value of the four--sided Wilson loop perfectly matches the $\mc N=4$ SYM one at two loops \cite{Andree:2010na}.
Furthermore we found that the two--loop amplitude in $\mc N=2$ SCQCD does not respect the maximum  transcendentality principle.

The absence of duality and the lack of dual conformal invariance might give an insight into the possible presence of integrable structures in $\mc N=2$ SCQCD. Indeed, a spin chain picture for the composite operators has been introduced and the properties of the dilatation operator have been studied \cite{Gadde:2010zi, Liendo:2011xb, Pomoni:2011jj}. As a result, even if the full $\mc N=2$ SCQCD model was not found to be integrable \cite{Gadde:2012rv}, it was suggested that the $SU(2,1|2)$ sector might be integrable at all loops \cite{Pomoni:2013poa} and it was argued that its integrable structure might be obtained from the one of $\mathcal{N}=4$ SYM by simply replacing the gauge coupling with an effective coupling \cite{Mitev:2014yba}. 

It would be then interesting to explore the consequences of the absence of duality and the lack of dual conformal invariance of our result \eqref{result} on the integrability of the $SU(2,1|2)$ sector. In fact, this sector of the theory is obtained by restricting to operators built with selected fields in the $\mc N=2$ vector multiplet, in a similar way as we only consider amplitudes with external adjoint particles.  In this direction it would be important to understand if it is possible to apply the knowledge of amplitudes in $\mc N=2$ SCQCD to compute off--shell quantities, and in particular the dilatation operator. Recently some first attempts to find this connection in $\mc N=4$ SYM were done in \cite{Zwiebel:2011bx, Wilhelm:2014qua, Brandhuber:2014pta, Brandhuber:2015boa}, applying methods originally devised for computing MHV amplitudes to the derivation of the one--loop dilatation operator and of the anomalous dimensions at higher loops \cite{Nandan:2014oga}. Such techniques might help to connect our on--shell results to the spin chain picture  also in the $\mc N=2$ SCQCD case.

\section{Acknowledgements}

This work has been supported in part by INFN and MPNS-COST Action MP1210  ``The String Theory Universe".

\end{document}